\documentclass[english, a4]{article}
\usepackage[T1]{fontenc}
\usepackage[utf8]{inputenc}
\usepackage{color}
\usepackage{cite}
\usepackage{amstext}
\usepackage{graphicx}
\usepackage{amsmath}
\usepackage{subfigure}

\makeatletter

\providecommand{\tabularnewline}{\\}

\makeatother

\usepackage{babel}
\begin{document}

\title{Light Cone Distribution Amplitudes  of Excited  P-Wave Heavy Quarkonia at the Leading Twist}


\author{M. A. Olpak\footnote{olpak@metu.edu.tr (corresponding author)} \and A. Ozpineci \footnote{ozpineci@metu.edu.tr}\and V. Tanriverdi \footnote{tanriverdivedat@googlemail.com} \\ Middle East Technical University, Department of Physics, Ankara, Turkey}
\maketitle

\begin{abstract}
Leading twist light cone distribution amplitudes (LCDAs) are key ingredients in calculating various hadronic amplitudes using light cone QCD sum rules. 
This work concentrates on calculating the leading twist LCDAs of P-wave heavy quarkonia. Quark model wavefunctions for the ground, first and second excited states of P-wave charmonia and bottomonia have been calculated, and are used for calculating the relevant LCDAs and leptonic decay constants. 
\end{abstract}

\section{Introduction}

Understanding hadron structure and spectrum has been one of the major issues in high energy physics for over half a century. Various models have been studied for this purpose up to now, either treating hadrons as fundamental (structureless) particles, or composite systems. Today, it is mostly believed that quantum chromodynamics (QCD) is the correct model of the fundamental constituents of hadrons \cite{pgw, colangelo}. However, it appears to be dramatically difficult to explain hadron structure and spectrum relying solely on QCD, and phenomenological models (such as non-relativistic or relativized quark models) are still relevant for studies in hadronic physics (e.g. see \cite{voloshin}), although the connection between such models and QCD has not been rigorously established up to now. 

The main difficulty in understanding hadron structure is in revealing the source of the confinement phenomenon on theoretical grounds. However, the phenomenon is continuously being demonstrated experimentally (no free quarks or glouns have been detected up to now) and has to be taken into account for understanding properties of hadrons. This issue motivates the use of potential models, which also involve a "confinement potential"  \cite{godfrey}. In the seminal paper \cite{godfrey}, a ``relativized'' quark model motivated by QCD  is constructed and the spectrum  and various transitions of mesons are calculated. 

Due to its non-perturbative nature, non-perturbative methods are necessary to study the hadronic spectrum. One of these methods for analyzing the spectrum and interactions of hadrons is provided by QCD sum rules \cite{SVZ} or its improved light cone QCD sum rules \cite{colangelo, voloshin, balitsky}. However, this method provides reliable  information concerning only states that are not radially excited, while there are many known radially excited states in the hadron spectrum. There had been some efforts to study the radially excited states in the literature (see e.g. \cite{Hohler:2012xd} and \cite{DiSalvo:1998hz}). Other than potential models, the most promising method for studying excited states (as well as all other properties of the hadron spectrum) is lattice QCD, which also has an extensive literature \cite{voloshin}.

All methods concentrate on calculating physically observable quantities related to hadrons and hadron interactions, though may be regarding part of those as inputs (e. g. a number of hadron masses) for the calculations. Hadron interactions constitute part of the observables involving hadrons. In light cone QCD sum rules, these interactions are expressed in terms of light-cone distribution amplitudes (LCDAs) \cite{colangelo, brodsky, lepage, brodsky_e, brodsky_q, brodsky_kitap, yang, braguta_mass, braguta_eta, braguta_2s, braguta_jpsi, braguta_p, ozpineci, braun, braun_l}. 
Hence, it is of crucial importance to be able to calculate these LCDAs for the hadrons. One proposed  way to obtain the leading twist LCDAs is to  use the non-relativistic quark model wave functions obtained through some potential quark models \cite{braguta_p, braguta_2s, hwang}. One advantage of this approach is that it also allows one to obtain the LCDAs of the radially excited states \cite{braguta_2s}.

As more and more heavier quarkonia are being discovered in experiments, the question of radial excitation attracts attention. Especially excitations above open flavor thresholds present a puzzle for the potential quark model calculations. There are indications that some of these quarkonia, close to or above thresholds, contain both a molecular component, and a quarkonium component \cite{voloshin, takizawa, cincioglu}. Although these quarkonium components themselves are not directly observable in nature, to study these mixed quarkonium-molecular systems, it is necessary to know the couplings of the quarkonium component to the molecular component \cite{takizawa, cincioglu}. LCDAs obtained through wave functions calculated using potential quark models can be a window to study such systems. 

In light cone QCD sum rules, to study the coupling of the hadron $H$ to those which can be created by the currents $j$ and $j'$, a correlation correlation function of the following form
\begin{equation}
F(p,q)=i\int d^{4}x e^{- i q \cdot x}\langle 0 \vert T{j'(x) j^{\dagger }(0)} \vert H(p) \rangle
\label{correl}
\end{equation}
is analyzed \cite{colangelo, brodsky, lepage, ozpineci, braun, braun_l}. In Eq. (\ref{correl}), the hadron $H(p)$ is on-shell. Such correlation functions reduce to expressions involving LCDAs once an expansion around $x^{2}=0$ is performed \cite{colangelo, brodsky, lepage, ozpineci, braun, braun_l}. Such an expansion allows one to perform a partial summation of the operators appearing in a usual $x=0$ expansion in terms of their twist, which is defined as the difference between the dimension and spin of an operator \cite{colangelo}. Although sum rules approach can be used for calculating the LCDAs as well (e. g. \cite{braguta_mass, braguta_eta, braguta_2s, braguta_jpsi, braguta_p}), LCDAs corresponding to excited states cannot be calculated using this method. Relating the LCDAs to non-relativistic potential models, circumvents this difficulty.

The connection between wave functions (calculated by any means, not necessarily using a potential model) and LCDAs has already been studied (e. g. see \cite{hwang}). In \cite{hwang}, ground state P-wave quarkonium wave functions obtained using a variational wave function have been used to obtain the LCDAs. In \cite{braguta_2s}, excited S-wave charmonia are studied. However, LCDAs corresponding to the excited P-wave quarkonium states are still to be discussed. In this work, relations between the quark model wave functions and leading twist LCDAs obtained in \cite{hwang} are used. The quark model wavefunctions are obtained by explicitely solving the model presented in \cite{godfrey}.

This work is organized as follows. In section 2, definitions of light cone coordinates and main results obtained in \cite{hwang} are summarized. In this section, the quark model of \cite{godfrey} is also presented shortly. Section 3 is devoted to the numerical analysis of our results, and describing model functions for the LCDAs. Finally,  we conclude our work in section 4.


\section{Leading Twist Light-cone distribution amplitudes for P-Wave quarkonia}   

The components of some 4-vector $k$ in light cone coordinates are defined as\cite{brodsky, lepage, brodsky_e, brodsky_q, brodsky_kitap, hwang}:

\begin{equation}
k^{\pm }\equiv k^{0}\pm k^{3}, \; \vec{k}_{\perp}=(k^{1},k^{2}).
\end{equation}
For a system of particles having total momentum $P$, one can define the light-cone momentum fractions $u_{i}$ as:

\begin{equation}
u_{i}\equiv k^{+} _{i}/P^{+}
\end{equation}
where $k_{i}$ are the momenta of the constituents. The LCDAs, as functions of these light-cone momentum fractions, are obtained by integrating over the transverse momenta. Defined in such a way, they are frame-independent quantities \cite{brodsky, lepage, brodsky_e, brodsky_q, brodsky_kitap, hwang}. 

For the wave functions, one considers the quark-gluon Fock states carrying the quantum numbers of the hadron, and calculates the LCWFs of each state contributing to the hadron state \cite{brodsky, lepage, brodsky_e, brodsky_q, brodsky_kitap, hwang}:

\begin{align}
|M(P;^{2S+1}L_{J_{z}},J_{z})>=& \sum_{Fock\, states}\int\left[\prod_{i}\frac{dk_{i}^{+}d^{2}k_{\perp,i}}{2(2\pi)^{3}}\right]2(2\pi)^{3}\delta^{(3)}\left(\tilde{P}-\sum_{i}\tilde{k}_{i}\right) \nonumber \\
& \times \sum_{\lambda_{i}}\Psi_{LS}^{JJ_{z}}(\tilde{k}_{i},\lambda_{i})|relevant\; Fock\; state>.
\end{align}                
where $\tilde{k}=(k^{+},\vec{k}_{\perp})$ and $\Psi_{LS}^{JJ_{z}}(\tilde{k}_{i})$ are the light cone wave functions corresponding the the given hadron quantum numbers and relevant Fock states. When one wishes to calculate hadronic couplings, one encounters matrix elements of the form \cite{colangelo, yang, braguta_mass, braguta_eta, braguta_2s, braguta_jpsi, braguta_p} $\left\langle q'\bar{q}(p,\epsilon)\left|\bar{q}'(x)\Gamma q(y)\right|0\right\rangle$. For light like separations $x-y$, this matrix element can be written as
\begin{equation}
\left\langle q'\bar{q}(p,\epsilon)\left|\bar{q}'(x)\Gamma q(y)\right|0\right\rangle =-f_{q'\bar{q}}\int_{0}^{1}du\exp(iup\cdot x+i \bar u p\cdot y)
\Phi (u) V[\Gamma ],
\end{equation}
where $\Phi (u)$ is the leading twist distribution amplitude of the $q'\bar{q}$ system and $V[\Gamma ]$ represents the Lorentz structure related to the Dirac matrix structure $\Gamma $ and possible other factors. Through the hadron states, wavefunctions enter the calculation and one can obtain corresponding LCDAs in terms of the relevant wavefunctions. 

In \cite{hwang}, a detailed analysis is presented on how to relate the wave functions to the LCDAs. For completeness, main points in their derivation is presented below.
Leading twist distribution amplitudes of P-wave heavy quarkonia are extracted from the following matrix elements:
\begin{align}
\left\langle 0\left|\bar{q}(-z)\gamma^{\mu}q(z)\right|S(P)\right\rangle |_{z^{2}=0} &=f_{S}P^{\mu}\intop_{0}^{1}du e^{i  \xi Pz} \phi_{S}(u),
\\
\left\langle 0\left|\bar{q}(-z)\gamma^{\mu}q(z)\right|A(P,\epsilon_{\lambda=0})\right\rangle |_{z^{2}=0} &=if_{A}M_{A}\epsilon^{\mu}\intop_{0}^{1}due^{i  \xi Pz} \phi_{A\parallel}(u),
\\
\frac{1}{2}\left\langle 0\left|\bar{q}(-z)(\gamma^{\mu}z\cdot\hat{D}+\gamma^{\nu}z_{\nu}\hat{D}^{\mu})q(z)\right|T(P,\epsilon_{\lambda=0})\right\rangle |_{z^{2}=0} &=f_{T}M_{T}^{2}\epsilon^{\mu\nu}z_{\nu}\intop_{0}^{1}due^{i  \xi Pz} (1-2u)\phi_{T\parallel}(u),
\\
\left\langle 0\left|\bar{q}(-z)\sigma^{\mu\nu}z_{\nu}\epsilon_{\perp\mu}\gamma_{5}q(z)\right|A(P,\epsilon_{\lambda=\pm1})\right\rangle |_{z^{2}=0} &=f_{A}^{\perp}\intop_{0}^{1}due^{i  \xi Pz} \phi_{A\perp}(u)\left(\epsilon\cdot\epsilon_{\perp}\right)\left(P\cdot z\right),
 \\ 
\left\langle 0\left|\bar{q}(-z)\sigma^{\mu\nu}z_{\nu}\epsilon_{\perp\mu\rho}z^{\rho}q(z)\right|T(P,\epsilon_{\lambda=\pm1})\right\rangle |_{z^{2}=0} &=if_{T}^{\perp}M_{T}\intop_{0}^{1}due^{i  \xi Pz} \phi_{T\perp}(u)\left(\epsilon^{\mu\nu}z_{\nu}\epsilon_{\perp\mu\rho}z^{\rho}\right),
\end{align}
where $z$ is the spacetime seperation between the quark and the anti-quark, $\xi=1-2u$, $\epsilon^{\mu}$ and $\epsilon^{\mu \nu}$ are the polarization vector and tensor of the relevant mesons, $P$ is the four-momentum, $M$ and $f$ are the mass and decay constant of the relevant mesons, and $\hat{D}$ is the gauge covariant derivative. The abbreviations $S$, $A$ and $T$ correspond to scalar, axial-vector and tensor, respectively. Using the C-parity, it can be shown that the distribution amplitudes should have definite symmetry properties under reflections through $u=\frac12$. The P-wave scalar and tensor mesons have positive C-parities, and hence, their DAs are odd under the exchange of $u \leftrightarrow \bar u$, where $\bar{u}=1-u$. The axial vector can be C-odd or C-even. For C-odd axial vector $\phi_{A\parallel}$ is odd and $\phi_{A\perp}$ is even, and for C-even axial vector, $\phi_{A\parallel}$ is even and $\phi_{A\perp}$ is odd.

Taking only the quark-antiquark component of the quarkonia, Leading twist distribution amplitudes are related to the quark model wave-function through \cite{hwang}:  
\begin{align}
\phi_{^{3}P_{0}}(u)&=\frac{\sqrt{2}}{f_{^{3}P_{0}}}\int\frac{d^{2}\kappa_{\perp}}{2(2\pi)^{3}}\frac{(1-2u)m_{q}}{\sqrt{u\bar u}}\varphi_{^{3}P_{0}}(u,\vec{\kappa}_{\perp}),
\label{phi3p0}
\\
\phi_{^{3}P_{1}\parallel}(u)&=\frac{2\sqrt{3}}{f_{^{3}P_{1}\parallel}}\int\frac{d^{2}\kappa_{\perp}}{2(2\pi)^{3}}\frac{\kappa_{\perp}^{2}}{\sqrt{u\bar u}M_{0}(m_{q},u,\vec{\kappa}_{\perp})}\varphi_{^{3}P_{1}\parallel}(u,\vec{\kappa}_{\perp}),
\\
\phi_{^{1}P_{1}\parallel}(u)&=\frac{\sqrt{6}}{f_{^{1}P_{1}\parallel}}\int\frac{d^{2}\kappa_{\perp}}{2(2\pi)^{3}}\frac{(1-2u)m_{q}}{\sqrt{u\bar u}}\varphi_{^{1}P_{1}\parallel}(u,\vec{\kappa}_{\perp}),
\\
\phi_{^{3}P_{2}\parallel}(u)&=\frac{\sqrt{6}}{f_{^{3}P_{2}\parallel}}\int\frac{d^{2}\kappa_{\perp}}{2(2\pi)^{3}}\frac{(1-2u)}{\sqrt{u\bar u}}\left[M_{0}(m_q,u,\vec{\kappa}_{\perp})-m_{q}-\frac{\kappa_{\perp}^{2}}{M_{0}(u,\vec{\kappa}_{\perp})+2m_{q}}\right]\varphi_{^{3}P_{2}\parallel}(u,\vec{\kappa}_{\perp}),
\\
\phi_{^{3}P_{1}\perp}(u)&=\frac{\sqrt{3}}{f_{^{3}P_{1}\perp}}\int\frac{d^{2}\kappa_{\perp}}{2(2\pi)^{3}}\frac{(1-2u)m_{q}}{\sqrt{u\bar u}}\varphi_{^{3}P_{1}\perp}(u,\vec{\kappa}_{\perp}),
\\
\phi_{^{1}P_{1}\perp}(u)&=\frac{\sqrt{6}}{f_{^{1}P_{1}\perp}}\int\frac{d^{2}\kappa_{\perp}}{2(2\pi)^{3}}\frac{\kappa_{\perp}^{2}}{\sqrt{u\bar u}M_{0}(m_{q},u,\vec{\kappa}_{\perp})}\varphi_{^{1}P_{1}\perp}(u,\vec{\kappa}_{\perp}),
\\
\phi_{^{3}P_{2}\perp}(u)&=\frac{\sqrt{6}}{f_{^{3}P_{2}\perp}}\int\frac{d^{2}\kappa_{\perp}}{2(2\pi)^{3}}\frac{(1-2u)}{\sqrt{u\bar u}}\left[m_{q}+\frac{2\kappa_{\perp}^{2}}{M_{0}(m_{q},u,\vec{\kappa}_{\perp})+2m_{q}}\right]\varphi_{^{3}P_{2}\perp}(u,\vec{\kappa}_{\perp}),
\label{phi3p2}
\end{align}
where $m_{q}$ is the mass of the quark, $\varphi_{\cal M}(m_{q},u,\vec{\kappa}_{\perp})$ are the wave function of the state ${\cal M}$,  and 
\begin{equation}
M^{2} _{0}=\frac{m^{2} _{q}+\kappa^{2} _{\perp}}{u} + \frac{m^{2} _{q}+\kappa^{2} _{\perp}}{\bar u}.
\end{equation}
In Eqs. (\ref{phi3p0})-(\ref{phi3p2}), spectral notation is used, such that the scalar meson $S$ is the $^3P_0$ state, the axial vector mesons are the $^3P_1$ and $^1P_1$ states, and the tensor meson is the $^3P_2$ state. The leptonic decay constants can be obtained through the normalization condition for the distribution amplitudes:
\begin{equation}
\intop_{0}^{1}du\phi_{even}(u)=1,\;\intop_{0}^{1}du(1-2u)\phi_{odd}(u)=1.
\end{equation}
where $\phi_{even(odd)}$ is a distribution amplitude that is even(odd) with respect to $u=\frac12$.

The functions $\varphi_{\cal M}$ used in Eqs. (\ref{phi3p0})-(\ref{phi3p2}) can be related to the quark model wave functions as follows. Let
\begin{equation}
\varphi _{\cal M}(u, \kappa _{\perp})= \varphi _{p}(u, \kappa _{\perp})\kappa _{L_{3}}(u, \kappa _{\perp})
\end{equation}
where $\kappa _{L_{3}=\pm 1}=(\kappa_{1} \mp i \kappa_{2})/\sqrt{2}$ and $\kappa _{L_{3}=0}=\kappa_{3}(u, \kappa_{\perp})$. 

If $K(\vert \vec{\kappa }\vert)$ is the radial function calculated in terms of the standard Minkowski coordinates, the function $\varphi _{p}$ can be related to the function $K(\vert \vec{k}\vert)$ as 
\begin{equation}
\varphi _{p}(u, \kappa _{\perp})= A\times \sqrt{\frac{\partial \kappa _{z}}{\partial u} (u, \kappa _{\perp})}\frac{K(\vert \vec{\kappa }\vert (u, \kappa _{\perp}))}{\vert \vec{\kappa }\vert},
\end{equation}
where $\vert \vec \kappa\vert(u,\kappa_\perp)$ is the relative momentum of the quark and anti-quark, $A$ is the normalization constant that can be obtained using the normalization condition for the functions $\varphi_{\cal M}$:
\begin{equation}
\int\frac{du d^{2}\kappa _{\perp }}{(2\pi )^{3}} \varphi _{\cal M}(u, \kappa _{\perp}) \varphi ^{*} _{\cal M'}(u, \kappa _{\perp}) =\delta _{\cal M M'}.
\label{norm_varphi}
\end{equation}

Note that, for a given radial excitation quantum number, and ignoring any spin or angular momentum dependent potentials, all the states considered in this work have the same $\varphi_p$. Under this assumption, the following relations between the leptonic decay constants and LCDAs are expected \cite{hwang}:

\begin{equation}
\sqrt{3}f_{^{3}P_{0}}=f_{^{1}P_{1}\parallel}=\sqrt{2}f_{^{3}P_{1}\perp}\equiv f_{odd},\;\frac{f_{^{3}P_{1}\parallel}}{\sqrt{2}}=f_{^{1}P_{1}\perp}\equiv f_{even}.
\label{relations23}
\end{equation}
 
\begin{equation}
\phi_{^{3}P_{0}}=\phi_{^{1}P_{1}\parallel}=\phi_{^{3}P_{1}\perp}\equiv\phi_{odd},\;\phi_{^{3}P_{1}\parallel}=\phi_{^{1}P_{1}\perp}\equiv\phi_{even}.
\end{equation}

Upto this point, relations between quark model wave functions and LCDAs are discussed. As a result, once the quark model wave function for a state is  calculated, its leading twist distribution amplitudes can be obtained through Eqs. (\ref{phi3p0})-(\ref{phi3p2}). 
In this work, quark model wavefunctions calculated using the Godfrey-Isgur Hamiltonian have been used. 
%
The Hamiltonian presented in \cite{godfrey} can be written as: 

\begin{equation}
H=H_{0} + H^{conf.} _{ij} + H^{hyp.} _{ij} + H^{so} _{ij} 
\label{hamiltonian}
\end{equation} 
where $H_{0}$ is the relativistic kinetic energy, $H^{conf.} _{ij}$ is the confinement potential, $H^{hyp.} _{ij}$ is the hyperfine potential and $H^{so} _{ij}$ is the spin-orbit interaction. By construction, this Hamiltonian is written in the meson rest frame so that its eigenvalues correspond directly to meson masses. Eigenfunctions of the 3-dimensional simple harmonic oscillator are chosen as the basis in which this hamiltonian is to be diagonilized. In terms of this basis, the eigenstates can be written as:

\begin{eqnarray}
& \Psi^{QM}(\vec{\kappa};n,L,S,J,J_{z})= \\ \nonumber
 & \sum_{L_{z},S_{z}}  \mathtt{C}\times\left\langle L,L_{z};S,S_{z}|L,S,J,J_{z}\right\rangle \times \chi_{S,S_{z}}\times Y_{L,L_{z}}(\theta_{\kappa},\phi_{\kappa}) \\ \nonumber
 &\sum_{m=0}^{N}h_{nm}\sqrt{2\times\frac{2}{\nu^{3}}\frac{m!}{\Gamma(m+L+\frac{1}{2})\left[2(L+m)+1\right]}}\left(\nu\kappa\right)^{L}\exp\left[-\frac{\kappa^{2}}{2\nu^{2}}\right]\mathtt{L}_{m}^{l+1/2}(\frac{\kappa^{2}}{\nu^{2}}),
\end{eqnarray}
where $\vec{\kappa}$ is the relative momentum of the quarks, $\mathtt{L}_{m}^{l+1/2}(\frac{\kappa^{2}}{\nu^{2}})$ are Laguerre polynomials, $Y_{L,L_{z}}(\theta_{\kappa},\phi_{\kappa})$ are spherical harmonics in momentum space. $n$ is the radial quantum number, $\mathtt{C}=\frac{1}{\sqrt{3}}(R\bar{R}+B\bar{B}+G\bar{G})$ is the color part and $\chi_{S,S_{z}}$ is the spin part of the wavefunction. To make a numerical diagonalization of the Hamiltonian possible, the Hamiltonian matrix is truncated by keeping only the first $N=16$ states in the corresponding block specified by the conserved quantities of the system.
$h_{nm}$ are determined by diagonalizing the $N\times N$ Hamiltonian matrix. The parameter $\nu$ appearing in the chosen basis, parametrizes the frequency of the oscillator. Its value is determined as to minimize the ground state energy in the corresponding Hamiltonian block.
Dressed $c$ quark and $b$ quark masses are taken to be $m_{c}=1628\, MeV,\; m_{b}=4977\, MeV$ in our calculations. Other parameters related to the quark model calculations can be found in \cite{godfrey}.

\begin{table}
\begin{center}
\begin{tabular}{|c|c|c|c|c|c|c|}
\hline 
Masses & \multicolumn{3}{c|}{$c\bar{c}$} & \multicolumn{3}{c|}{$b \bar{b}$} \tabularnewline
\hline 
\hline 
$M\,(GeV)\setminus n$           & $n=1$ & $n=2$ & $n=3$ & $n=1$ & $n=2$ & $n=3$\tabularnewline
\hline 
$M_{^{3}P_{0}}\, (\chi _{q0})$ & $3.37$ & $3.88$ & $4.30$ & $9.81$ & $10.2$ & $10.7$\tabularnewline
\hline 
$M_{^{3}P_{1}}\, (\chi _{q1})$ & $3.54$ & $3.97$ & $4.33$ & $9.89$ & $10.3$ & $10.6$\tabularnewline
\hline 
$M_{^{1}P_{1}}\, (h _{q})$       & $3.53$ & $3.96$ & $4.37$ & $9.88$ & $10.3$ & $10.6$\tabularnewline
\hline 
$M_{^{3}P_{2}}\, (\chi _{q2})$ & $3.54$ & $3.98$ & $4.34$ & $9.89$ & $10.3$ & $10.6$\tabularnewline
\hline 
\end{tabular}
\caption{Quark model masses calculated for the first three levels of charmonia and bottomonia}
\label{table:masses}
\end{center}
\end{table}

\begin{table}
\begin{center}
\begin{tabular}{|c|c|c|c|c|c|c|}
\hline 
Masses  & \multicolumn{3}{c|}{$c\bar{c}$} & \multicolumn{3}{c|}{$b \bar{b}$} \tabularnewline
\hline 
\hline 
$M\,(MeV)\setminus n$           & $n=1$       & $n=2$       & $n=3$ & $n=1$ & $n=2$ & $n=3$\tabularnewline
\hline 
$M_{^{3}P_{0}}\, (\chi _{q0})$ & $3414.75$ & $-$           & $-$       & $9859.44$ & $10232.5$ & $-$\tabularnewline
\hline 
$M_{^{3}P_{1}}\, (\chi _{q1})$ & $3510.66$ & $-$           & $-$       & $9892.78$ & $10255.46$ & $10512.1$\tabularnewline
\hline 
$M_{^{1}P_{1}}\, (h _{q})$       & $3525.38$ & $-$           & $-$       & $9899.3$   & $10259.8$ & $-$\tabularnewline
\hline 
$M_{^{3}P_{2}}\, (\chi _{q2})$ & $3556.20$ & $3927.2$ & $-$       & $9912.21$ & $10268.65$ & $-$\tabularnewline
\hline 
\end{tabular}
\caption{Masses of experimentally observed states in Particle Data Group listings \cite{beringer}.}
\label{table:masses_exp}
\end{center}
\end{table}

Within the above mentioned framework, the spectrum and the wave functions are obtained. 
The obtained masses for the first three levels are presented in Table \ref{table:masses}. In Table \ref{table:masses_exp}, we present the observed masses (if available) for the corresponding states. Comparing the two tables, it is observed that the model is quite successful in reproducing the observed masses (when available).

\section{Numerical Analysis and Model LCDAs}

Once the wave functions are obtained, the calculation of the LCDAs are straightforward. One issue that needs to be addressed is the contribution from the relativistic tails of the wave functions. Although the model of \cite{godfrey} is relativized by the use of relativistic kinetic energy expression, inclusion of retardation effects, and smearing, it is still questionable how reliable the model can describe relativistic tails. If the relativistic tail does not contribute to a result,  than we can claim that our results are safe from any relativistic ``contamination,'' whereas if the relativistic tail dominates, our results should be used with caution. To estimate the contributions of the relativistic tail, the $\vec \kappa_\perp$ integrals appearing in Eqs. (\ref{phi3p0})-(\ref{phi3p2}) are cut at a cutoff $\vert \vec \kappa_\perp \vert \le \Lambda$. The results are evaluated both at $\Lambda=\infty$ and at $\Lambda=m_q$ ($q=c$ for the charmonia and $q=b$ for the bottomonia).

 On practical grounds, it is also desirable to express the LCDAs in terms of a few parameters which can be easily tabulated and used.
In \cite{braguta_p}, the following expressions for the LCDAs have been motivated using sum rules techniques (for $\xi $-odd and $\xi $-even LCDAs respectively):

\begin{align}
\phi_{odd} (\xi ) =& c(\beta )(1-\xi ^2)\xi \exp[-\frac{\beta }{(1-\xi ^2)}] \nonumber \\
\phi_{even} (\xi ) =& -\intop _{-1} ^{\xi} dt \phi_{odd} (t)=\frac{c(\beta )}{2}(1-\xi ^2)^2 E_{3} (\frac{\beta}{(1-\xi ^2)}) 
\end{align}
where $E_{3} (x)\equiv \intop _{1} ^{\infty } dt \frac{e^{-x}}{t^3}$, and the parameters $c$ and $\beta $ are to be fitted to the LCDAs. 
In this work, we generalize this model to the excited states as well.
The models for even LCDAs are chosen to be:

\begin{align}
n=1: \quad  \psi (\xi ) =& a (1-\xi ^2)^2 \left( E_{3} [\frac{\beta}{(1-\xi ^2)}] + b \exp[-\frac{\xi ^2}{c}]\right) \nonumber\\
n=2,\,3: \quad  \psi (\xi ) =& a \left\lbrace \frac{1}{1+\frac{(\xi ^2 - \xi _{0} ^2)^2}{\sigma ^{2}}} + b \exp[-\frac{\xi ^2}{c}]\right\rbrace \exp[-\frac{\beta }{(1-\xi ^2)}],
\end{align} 
and for odd LCDAs:
\begin{align}
n=1: \quad  \phi (\xi ) =& a \xi (1-\xi ^2)\lbrace \exp[-\frac{\beta }{(1-\xi ^2)}] + b \exp[-\frac{\xi ^2}{c}] \rbrace \nonumber \\
n=2,\, 3: \quad  \phi (\xi ) =& -\frac{d}{d \xi}\left\lbrace a \left[ \frac{1}{1+\frac{(\xi ^2 - \xi _{0} ^2)^2}{\sigma ^{2}}} + b \exp[-\frac{\xi ^2}{c}]  \right]  \exp[-\frac{\beta }{(1-\xi ^2)}]\right\rbrace  .
\end{align} 


The results for the relevant leptonic decay constants are presented in Tables \ref{DC1}-\ref{DC3}.  In the tables, both leptonic decay constans, and  the leptonic decay constant multiplied by the coefficient, when not equal to one, in the relations presented in Eq. (\ref{relations23}) are presented in order to facility their comparison. 


\begin{table}
\begin{center}
\caption{Decay constants $f_{^{1}P_{1}},f_{^{3}P_{0}},f_{^{3}P_{1\perp}}$ for relevant charmonia and bottomonia.}
\begin{tabular}{|c|c|c|c|c|c|c|}
\hline
$n\setminus f\,(GeV)$ & $f_{^{3}P_{0}}$ & $f_{^{3}P_{1}\perp}$ & $f_{^{1}P_{1}}$ & $\sqrt{3}f_{^{3}P_{0}}$ & $\sqrt{2}f_{^{3}P_{1}\perp}$  & $f_{odd}$ \tabularnewline
\hline 
\hline 
charmonia  &  \multicolumn{6}{|c|}{$\Lambda = \infty $}  \tabularnewline
\hline 
$n=1$ & $0.109$ & $0.0959$ & $0.142$ & $0.189$ & $0.136$ & $0.118$ \tabularnewline
\hline 
$n=2$ & $0.0801$ & $0.0881$ & $0.129$ & $0.139$ & $0.125$ & $0.105$ \tabularnewline
\hline 
$n=3$ & $0.0755$ & $0.0824$ & $0.133$ & $0.131$ & $0.117$ & $0.103$ \tabularnewline
\hline 
& \multicolumn{6}{|c|}{$\Lambda = m_{c} $} \tabularnewline
\hline
$n=1$ &$0.0916$ & $0.0875$ & $0.127$ & $0.159$ & $0.124$ & $0.105$ \tabularnewline
\hline
$n=2$ &$0.0588$ & $0.0741$ & $0.107$ & $0.102$ & $0.105$ & $0.0860$ \tabularnewline
\hline
 $n=3$&  $0.0459$ & $0.0615$ & $0.0946$ & $0.0795$ & $0.0870$ & $0.0735$ \tabularnewline 
\hline
bottomonia  & \multicolumn{6}{|c|}{$\Lambda = \infty $} \tabularnewline
\hline 
$n=1$ & $0.104$ & $0.0802$ & $0.119$ & $0.180$ & $0.113$ & $0.100$ \tabularnewline
\hline 
$n=2$ & $0.103$ & $0.0832$ & $0.124$ & $0.178$ & $0.118$ & $0.104$ \tabularnewline
\hline 
$n=3$ & $0.131$ & $0.0834$ & $0.143$ & $0.227$ & $0.118$ & $0.116$ \tabularnewline
\hline 
&   \multicolumn{6}{|c|}{$\Lambda = m_{b} $}  \tabularnewline
\hline
$n=1$ & $0.0972$ & $0.0794$ & $0.117$ & $0.168$ & $0.112$ & $0.0981$ \tabularnewline
\hline
$n=2$& $0.0976$ & $0.0822$ & $0.121$ & $0.169$ & $0.116$ & $0.101$ \tabularnewline
\hline
$n=3$& $0.118$ & $0.0820$ & $0.136$ & $0.204$ & $0.116$ & $0.0951$ \tabularnewline
\hline
\end{tabular}
\label{DC1}
\end{center}
\end{table}

\begin{table}
\begin{center}
\caption{Decay constants $f_{^{1}P_{1}\perp},f_{^{3}P_{1}}$ for relevant charmonia and bottomonia.}
\begin{tabular}{|c|c|c|c|c|c|c|c|c|}
\hline
$n\setminus f\,(GeV)$ & $f_{^{3}P_{1}}$ & $f_{^{1}P_{1}\perp}$ & $\frac{f_{^{3}P_{1}}}{\sqrt{2}}$ & $f_{even}$ & $f_{^{3}P_{1}}$ & $f_{^{1}P_{1}\perp}$ & $\frac{f_{^{3}P_{1}}}{\sqrt{2}}$ & $f_{even}$ \tabularnewline
\hline 
\hline 
charmonia  & \multicolumn{4}{|c|}{$\Lambda = \infty $}  &  \multicolumn{4}{|c|}{$\Lambda = m_{c} $}  \tabularnewline
\hline 
$n=1$ & $0.264$ & $0.199$ & $0.187$ & $0.232$ & $0.185$ & $0.133$ & $0.131$ & $0.159$ \tabularnewline
\hline 
$n=2$ & $0.279$ & $0.209$ & $0.197$ & $0.244$ & $0.143$ & $0.101$ & $0.101$ & $0.122 $\tabularnewline
\hline 
$n=3$ & $0.290$ & $0.246$ & $0.205$ & $0.268$ & $0.0852$ & $0.0595$ & $0.0603$ & $0.0724$ \tabularnewline
\hline
bottomonia & \multicolumn{4}{|c|}{$\Lambda = \infty $}  &  \multicolumn{4}{|c|}{$\Lambda = m_{b} $}  \tabularnewline
\hline 
$n=1$ & $0.182$ & $0.138$ & $0.129$ & $0.160$ & $0.173$ & $0.126$ & $0.122$ & $0.146$ \tabularnewline
\hline 
$n=2$ & $0.197$ & $0.148$ & $0.139$ & $0.173$ & $0.184$ & $0.135$ & $0.130$ & $0.156$ \tabularnewline
\hline 
$n=3$ & $0.204$ & $0.182$ & $0.144$ & $0.193$ & $0.187$ & $0.153$ & $0.132$ & $0.170$ \tabularnewline
\hline 
\end{tabular}
\label{DC2}
\end{center}
\end{table}

\begin{table}
\begin{center}
\caption{Decay constants $f_{^{3}P_{2}},f_{^{3}P_{2\perp}}$ for tensor charmonia and bottomonia.}
\begin{tabular}{|c|c|c|c|c|}
\hline 
$n\setminus f\,(GeV)$ & $f_{^{3}P_{2}}$ & $f_{^{3}P_{2}\perp}$ & $f_{^{3}P_{2}}$ & $f_{^{3}P_{2}\perp}$\tabularnewline
\hline 
\hline 
charmonia  &\multicolumn{2}{|c|}{$\Lambda =\infty $} &  \multicolumn{2}{|c|}{$\Lambda = m_{c}$}  \tabularnewline
\hline
$n=1$ & $0.198$ & $0.141$ & $0.177$ & $0.128$\tabularnewline
\hline 
$n=2$ & $0.229$ & $0.142$ & $0.189$ & $0.118$\tabularnewline
\hline 
$n=3$ & $0.245$ & $0.140$ & $0.182$ & $0.101$\tabularnewline
\hline
bottomonia &\multicolumn{2}{|c|}{$\Lambda =\infty $} &  \multicolumn{2}{|c|}{$\Lambda = m_{b}$}  \tabularnewline
\hline 
$n=1$ & $0.133$ & $0.113$ & $0.131$ & $0.112$\tabularnewline
\hline 
$n=2$ & $0.148$ & $0.121$ & $0.146$ & $0.119$\tabularnewline
\hline 
$n=3$ & $0.178$ & $0.137$ & $0.168$ & $0.131$\tabularnewline
\hline  
\end{tabular}
\label{DC3}
\end{center}
\end{table}

It is observed that the relations in Eq. (\ref{relations23}) are qualitatively satisfied. The largest deviation is observed in $f_{^3P_0}$, which can be as large as, e.g. $50\%$ for the $n=3$ bottomonium case. The values given in \cite{braguta_p} and \cite{hwang} and the results of this work agree in the order of magnitude of the numbers. The fact that there is no precise agreement in the decay constants stems from using different model functions to calculate the decay constants and LCDAs. It is also generally observed that the deviations from relations in Eq. (\ref{relations23}) are enhanced when $n$ increases, but is suppressed when the finite cut-off is used. Both are expected from spin-orbit effects as in both cases, either the system is already non-relativistic, or the relativistic effects are omitted by imposing a cut-off. 
The leptonic constant for the charmonium change significantly when a finite cut-off is used, whereas the change is not so significant for the bottomonium sector. The dependence of the cutoff also increases as $n$ increases in both sectors. However, spin weighted average $f_{odd}$ of charmonium also appears to be slightly affected with the use of the cut-off. 

In \cite{braguta_p} and \cite{hwang}, leptonic decay constants of only the ground states are analyzed ignoring spin effects. For comparison, we present  the spin averaged leptonic decay constant in Tables \ref{table:charmoniumcomparison} and \ref{table:bottomoniujcomparison}. As can be seen from the tables, results obtained in this work for $f_{odd}$ are larger by about $30\%$ from the results of \cite{hwang} in both sectors. For $f_{even}$, the discrepancy is even larger, and results obtained in this work are almost twice as large as the results of \cite{hwang}. In \cite{braguta_p}, only the result for $f_{even}$ for charmonium is available. The result of \cite{braguta_p} is in agreement with the result of this work.

\begin{table}\centering
\caption{Comparison of thespin averaged leptonic decays constant for the ground state charmonium with the results in the literature}

\begin{tabular}{|c|c|c|c|c|}
\hline
& \cite{hwang} & \cite{braguta_p} & this work ($\Lambda=\infty$) & this work ($\Lambda=m_c$) \\ \hline
$f_{odd}$(GeV) &  $0.088$ & $-$ & $0.118$ & $0.105$\\ \hline
$f_{even}$(GeV) & $0.109$ & $0.192$ & $0.232$ & $0.159$ \\ \hline
$f_{T\parallel}$(GeV) & $0.124$&$-$&$0.198$	&$0.177$\\ \hline
$f_{T\perp}$(GeV) & $0.098$ &$-$&$0.141$	&$0.128$\\ \hline
\end{tabular}
\label{table:charmoniumcomparison}
\end{table}

\begin{table}\centering
\caption{Comparison of thespin averaged leptonic decays constant for the ground state bottomonium with the results in the literature}
\begin{tabular}{|c|c|c|c|}
\hline
& \cite{hwang} &  this work ($\Lambda=\infty$) & this work ($\Lambda=m_b$) \\ \hline
$f_{odd}$(GeV) &  $0.067$		& $0.100$	& $0.098$ \\ \hline
$f_{even}$(GeV) & $0.072$		& $0.160$	& $0.146$\\ \hline
$f_{T\parallel}$(GeV) & $0.075$	&$0.133$	&$0.131$	\\ \hline
$f_{T\perp}$(GeV) & $0.069$		&$0.113$	&$0.112$\\ \hline
\end{tabular}
\label{table:bottomoniujcomparison}
\end{table}



\begin{figure}
\begin{center}
\begin{tabular}{cc}
$c\bar{c}$ & $b\bar{b}$ \\
\includegraphics[scale=1]{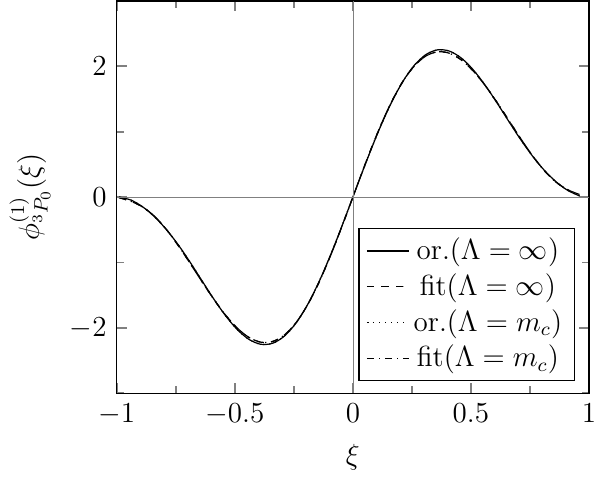} & \includegraphics[scale=1]{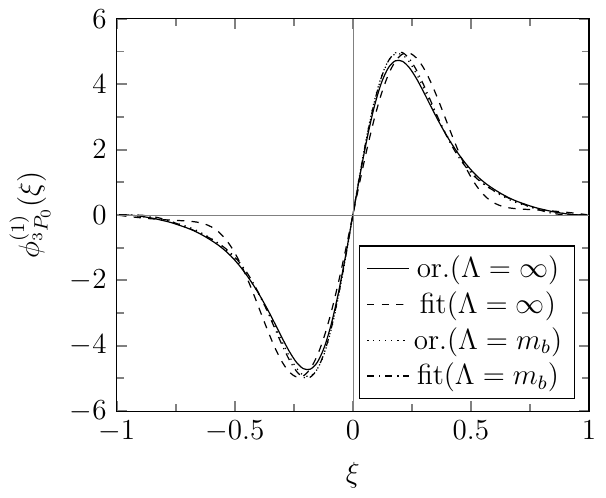} \\
\includegraphics[scale=1]{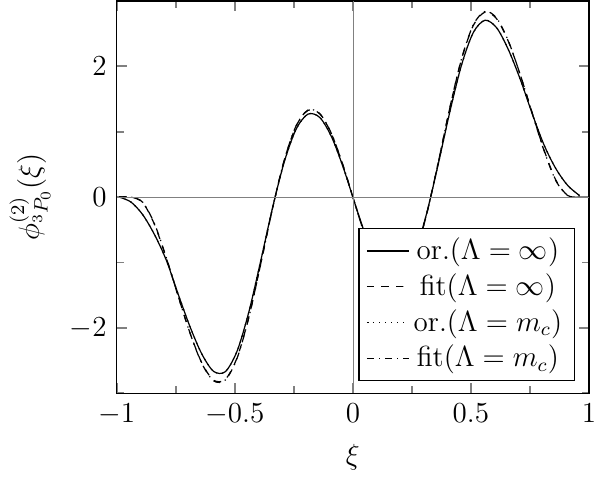} & \includegraphics[scale=1]{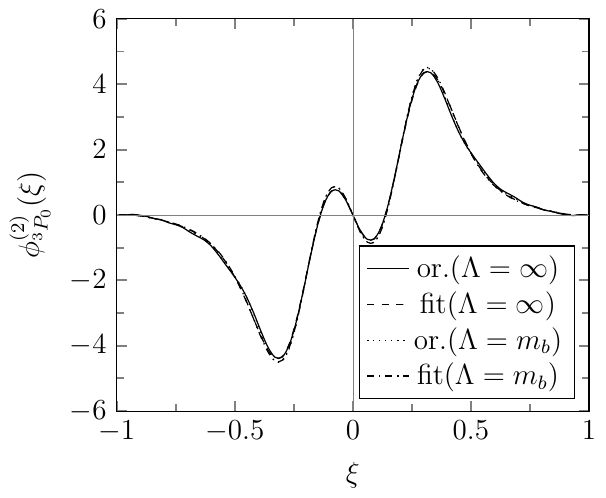} \\
\includegraphics[scale=1]{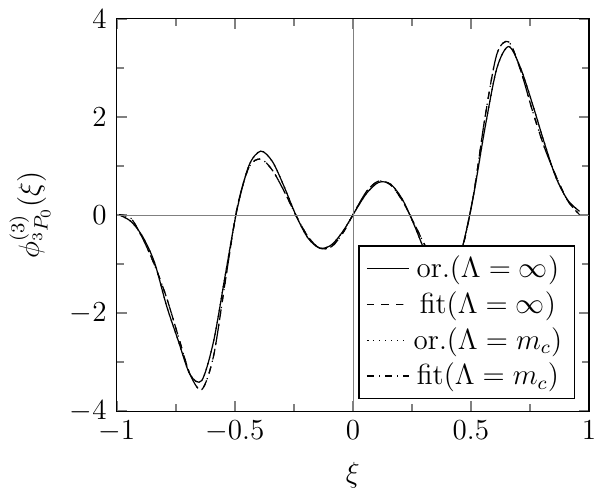} & \includegraphics[scale=1]{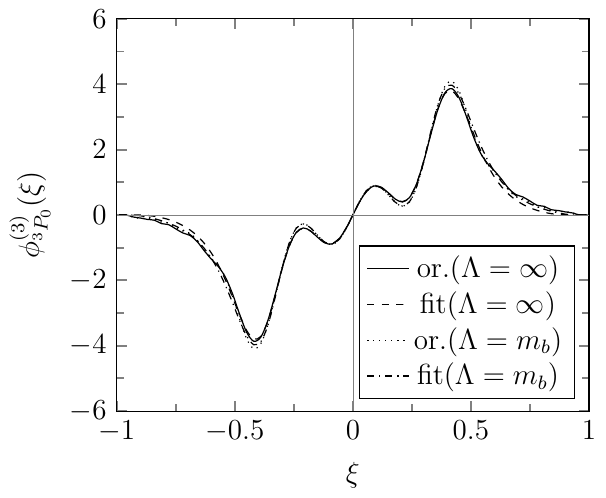} \\
\end{tabular}
\caption{LCDAs: $^{3} P_{0}$. Upper limit of $k_{\perp }$ integration is indicated in parantheses. "or." refers to the original function and "fit" refers to the fitted function. The radial quantum number $n$ is indicated in parantheses as superscript: $\phi ^{(n)}(u)$.}
\label{figilk}
\end{center}
\end{figure}

\begin{figure}
\begin{center}
\begin{tabular}{cc}
$c\bar{c}$ & $b\bar{b}$ \\
\includegraphics[scale=1]{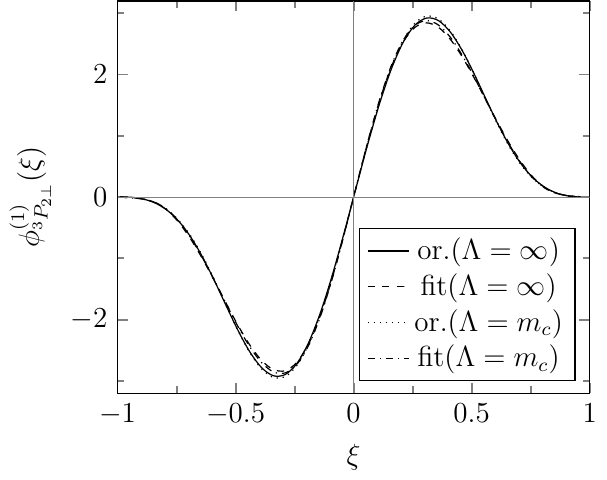} & \includegraphics[scale=1]{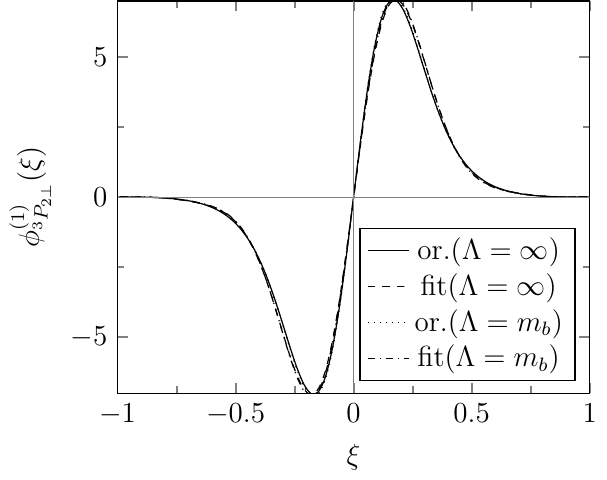} \\
\includegraphics[scale=1]{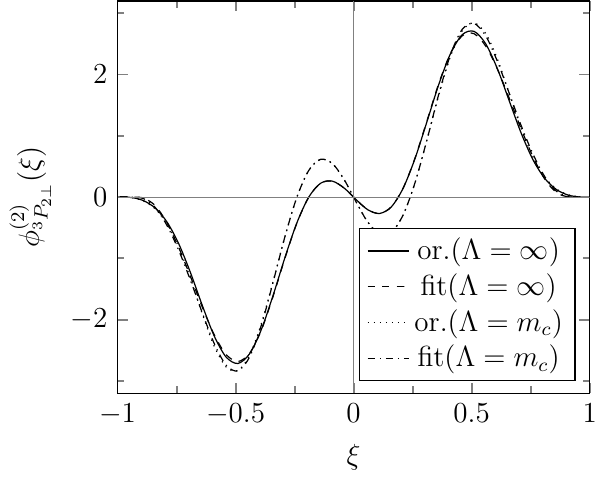} & \includegraphics[scale=1]{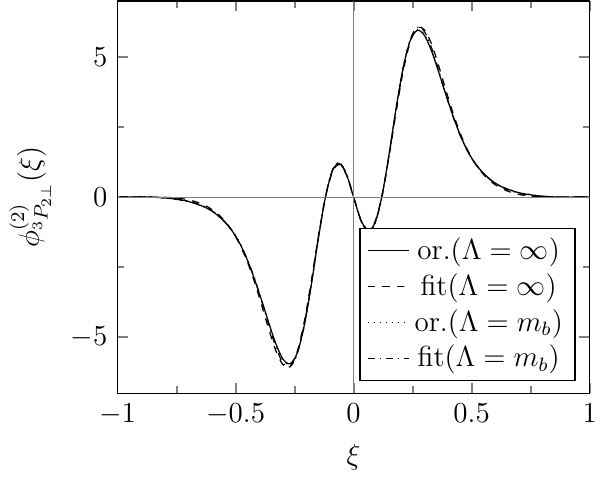} \\
\includegraphics[scale=1]{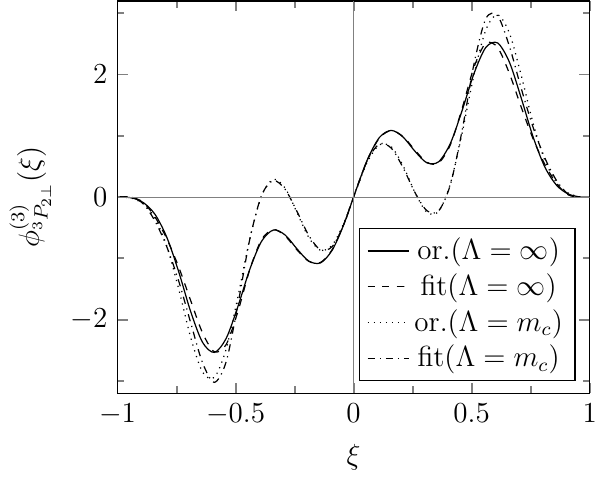} & \includegraphics[scale=1]{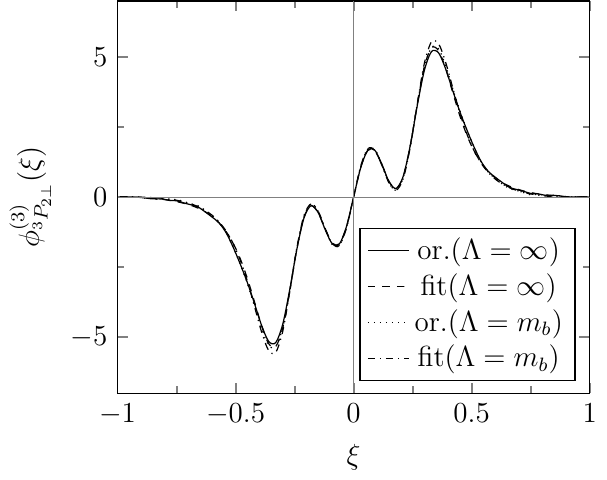} \\
\end{tabular}
\caption{LCDA plots as in Fig. (\ref{figilk}), but for $^{3} P_{2\perp}$ states.}
\label{figson}
\end{center}
\end{figure}

In Figs. (\ref{figilk}) - (\ref{figson}), LCDAs are depicted for the various states. In each of the plots, both the LCDAs obtained using Eqs. (\ref{phi3p0})-(\ref{phi3p2}) with $\Lambda=\infty$ and $\Lambda=m_q$, and also the fits to the LCDAs for both of $\Lambda$ values are shown.
The parameters used for each fit are presented in Tables (\ref{table:6})-(\ref{table:33}). As can be observed from the figures, the fits reliable reproduce the calculated LCDAs. Some general observations about the DAs are in order.
Odd DAs have $2n+1$ extrema for $u>0$ reflecting the nodal structure of the wave functions of the excited states. Even DAs have one (three) extrema when $n=1$ ($n=2$ or $n=3$). Some of the extrema for even DAs for $n=2$ and $n=3$ are converted into reflection points due to a nearby, larger extrema. As $n$ increases, some of the extrema move towards the $\xi =\pm 1$ region, which is the relativistic region. 
Similarly, the DAs for a given $n$ are localized closer to $\xi=0$ for bottomonium than for charmonium. This is again a reflection of the highly non-relativistic nature of the bottomonium system. As expected, another reflection of the non-relativistic nature of small $n$ and bottomonium system is the dependence on the cut-off. In general bottomonium systems and small $n$ systems are more non-relativistic compared to charmonium and large $n$ systems. 

\section{Conclusions}
In this work, the three lowest lying states of P-wave charmonia and bottomonia are considered.
Their LCDAs and decay constants have been calculated. 

It is observed the spin effects can be important in the determination of the leptonic decay constants. Also, leptonic decay constants and LCDAs receive larger contributions from relativistic effects for charmonium than for the bottomonium. The importance of the relativistic contributions becomes also larger as $n$ (the radial excitation quantum number) increases.

Also, the DAs of the bottomonium are closer to the $\xi=0$ region, than the charmonium DAs. Also, as $n$ increased, in both of the sectors DAs shift towards larger values of $\vert \xi\vert$, which is another indication that these systems become more relativistic.

For future usage, model functions have been fitted to the obtained LCDAs, so that the obtained DAs can be easily used in future works.


\section*{Acknowledgments}
This work is partially supported by TUBITAK under grant no 111T706.
\newpage
\newpage
\begin{figure}
\begin{center}
\begin{tabular}{cc}
$c\bar{c}$ & $b\bar{b}$ \\
\includegraphics[scale=1]{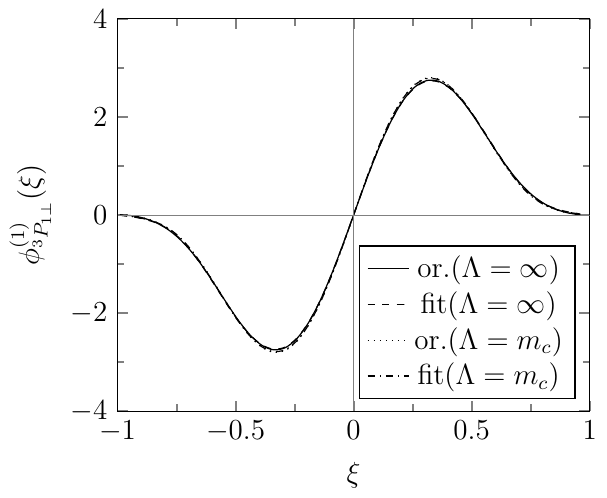} & \includegraphics[scale=1]{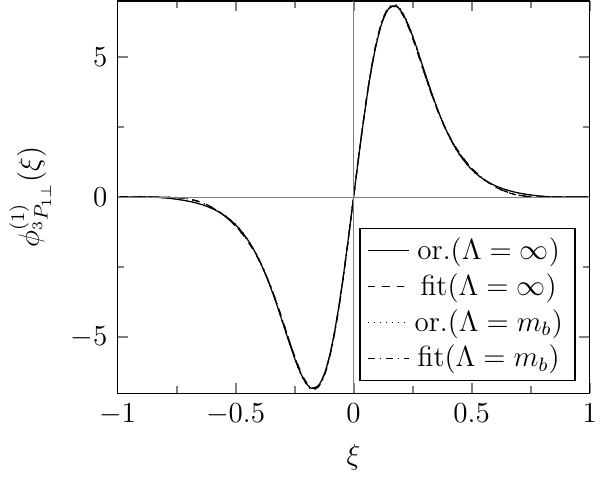} \\
\includegraphics[scale=1]{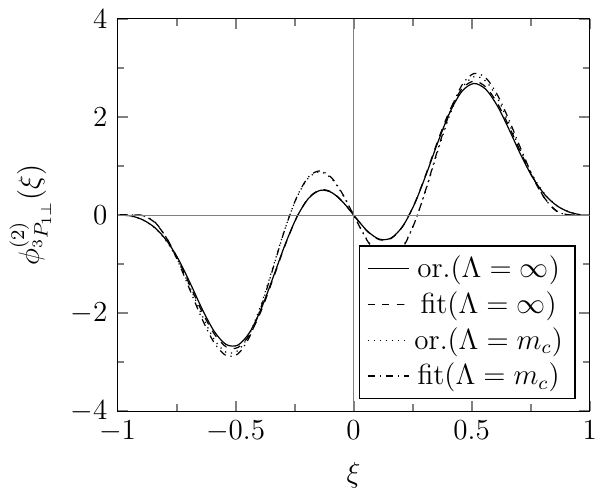} & \includegraphics[scale=1]{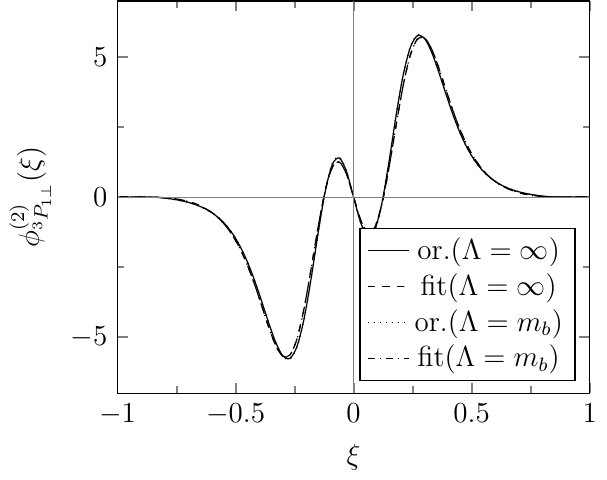} \\
\includegraphics[scale=1]{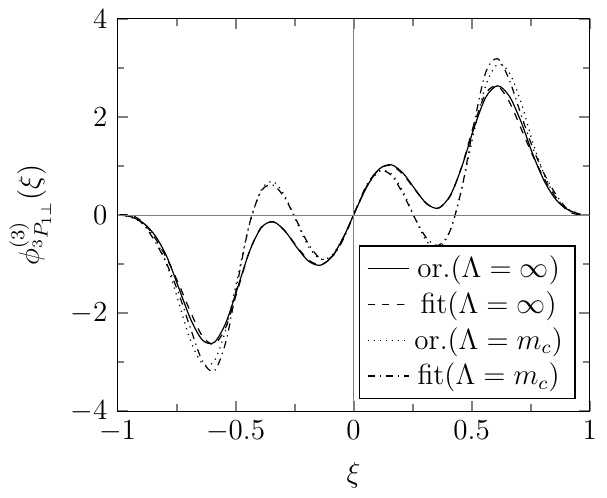} & \includegraphics[scale=1]{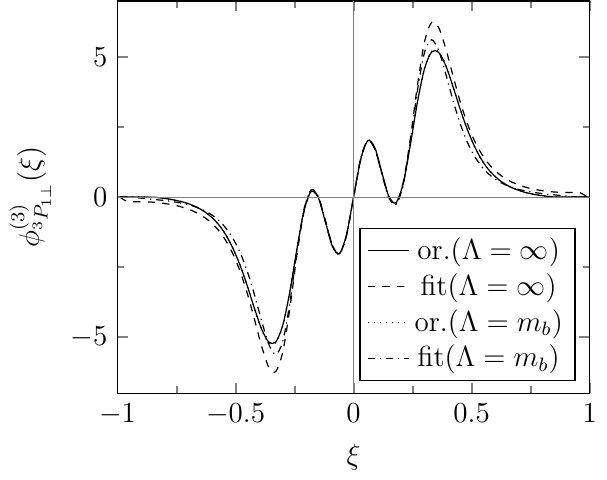} \\ 
\end{tabular}
\caption{LCDA plots as in Fig. (\ref{figilk}), but for $^{3} P_{1\perp}$ states.}
\end{center}
\end{figure}
\newpage
\begin{figure}
\begin{center}
\begin{tabular}{cc}
$c\bar{c}$ & $b\bar{b}$ \\
\includegraphics[scale=1]{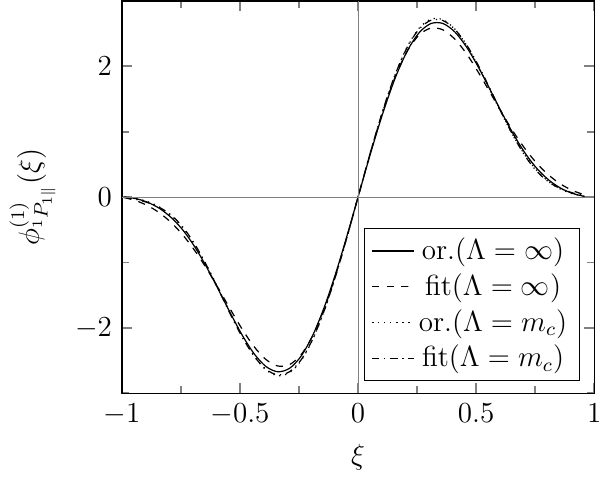} & \includegraphics[scale=1]{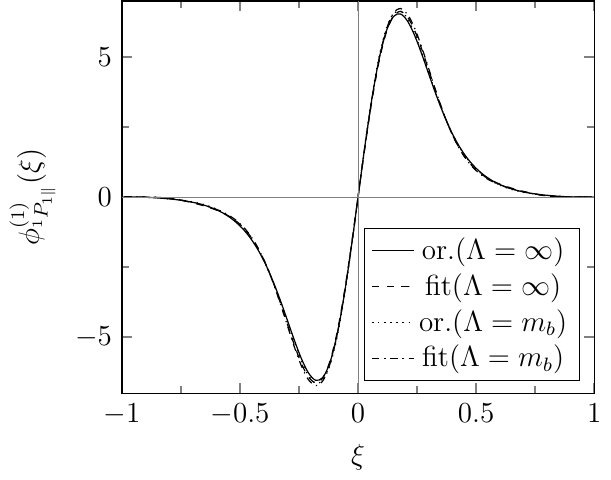} \\
\includegraphics[scale=1]{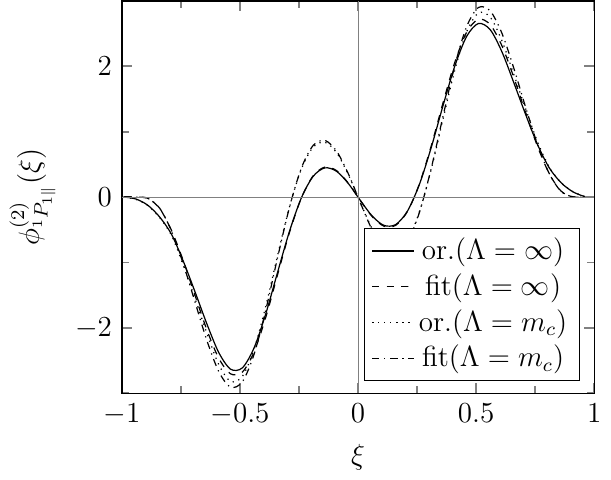} & \includegraphics[scale=1]{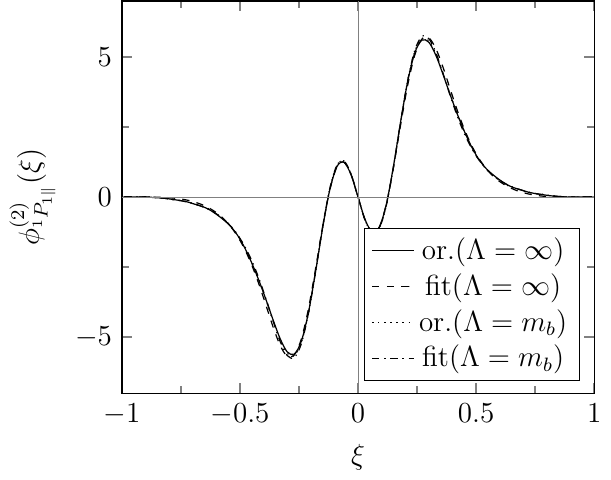} \\
\includegraphics[scale=1]{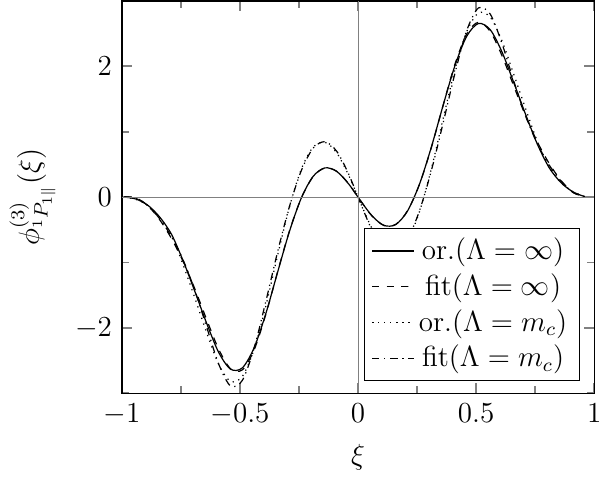} & \includegraphics[scale=1]{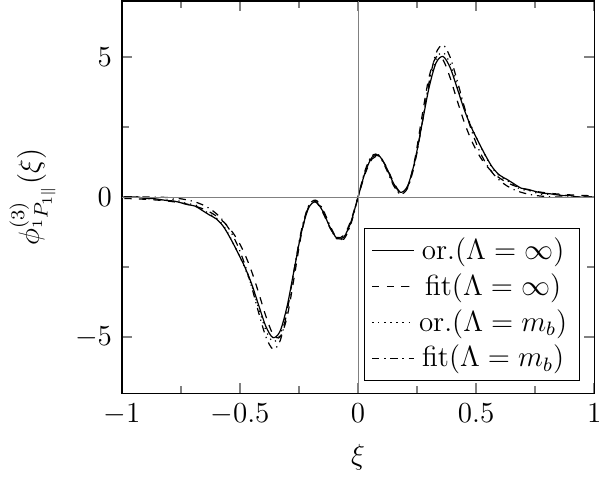} \\
\end{tabular}
\caption{LCDA plots as in Fig. (\ref{figilk}), but for $^{1} P_{1\parallel}$ states.}
\end{center}
\end{figure}
\newpage
\begin{figure}
\begin{center}
\begin{tabular}{cc}
$c\bar{c}$ & $b\bar{b}$ \\
\includegraphics[scale=1]{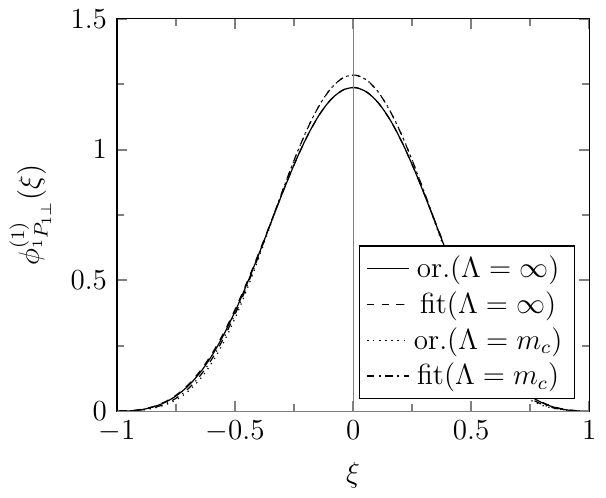} & \includegraphics[scale=1]{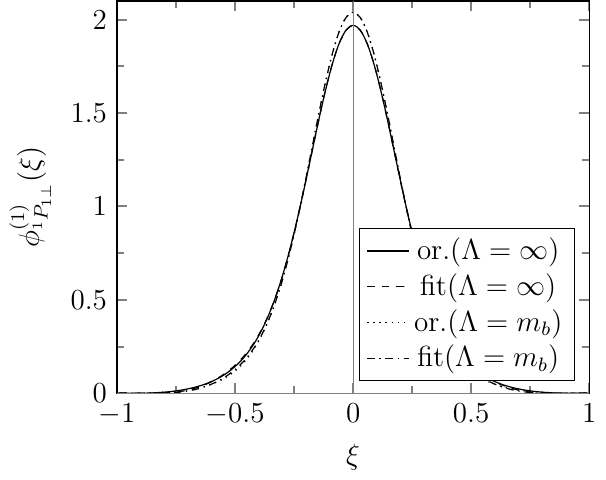} \\
\includegraphics[scale=1]{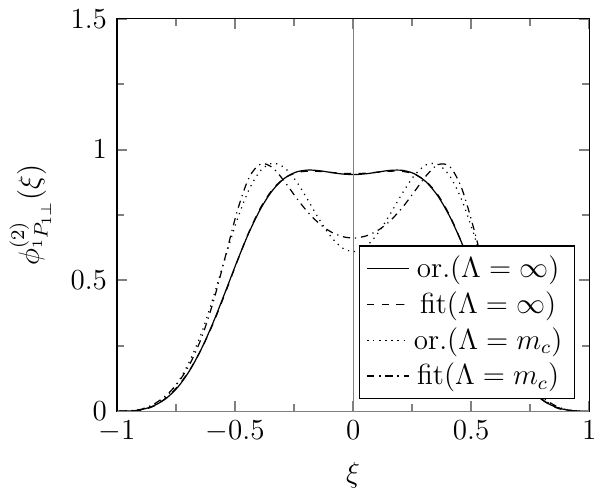} & \includegraphics[scale=1]{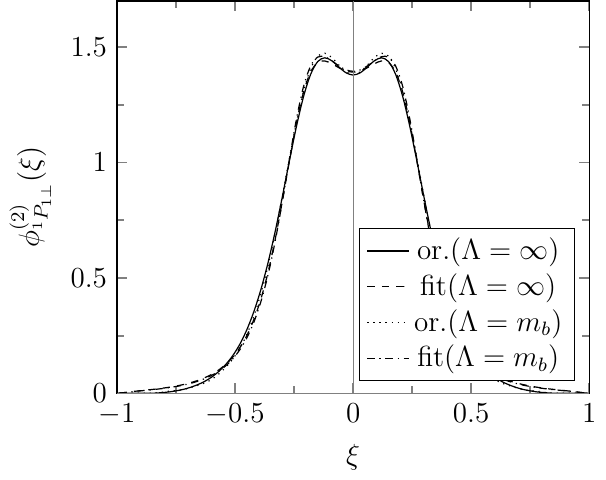} \\
\includegraphics[scale=1]{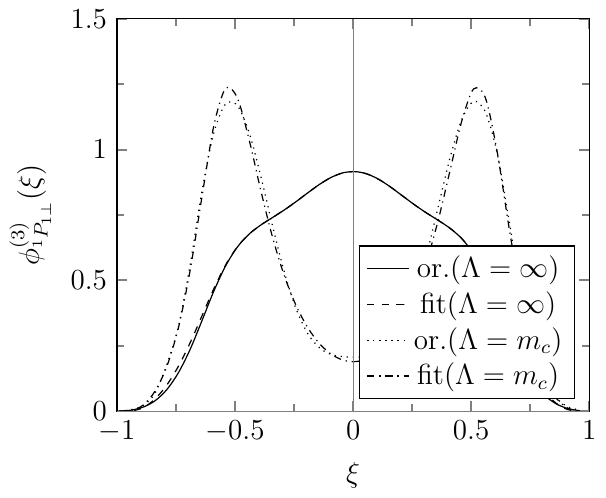} & \includegraphics[scale=1]{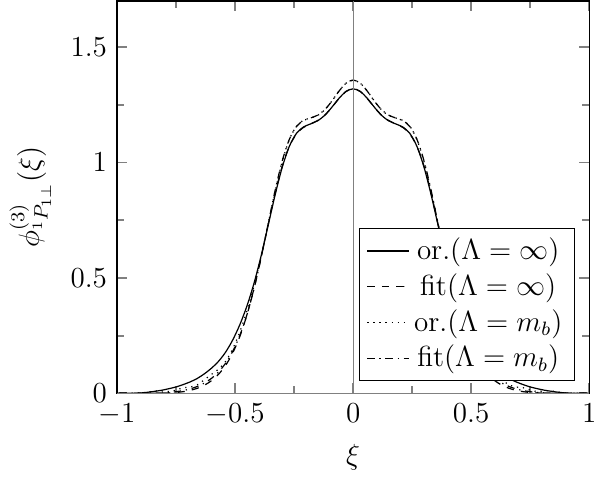} \\
\end{tabular}
\caption{LCDA plots as in Fig. (\ref{figilk}), but for $^{1} P_{1\perp}$ states.}
\end{center}
\end{figure}
\newpage
\begin{figure}
\begin{center}
\begin{tabular}{cc}
$c\bar{c}$ & $b\bar{b}$ \\
\includegraphics[scale=1]{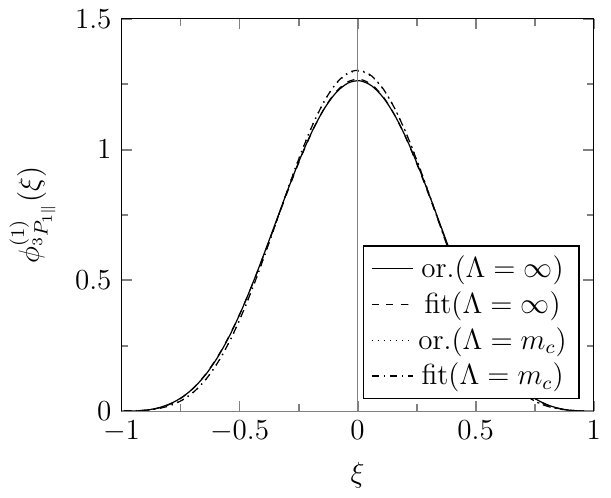} & \includegraphics[scale=1]{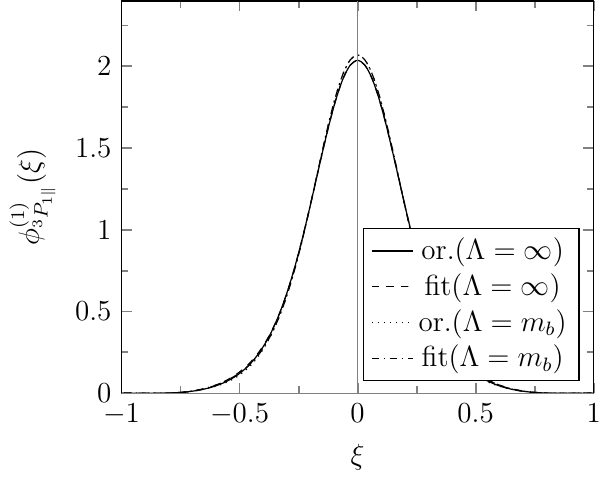} \\
\includegraphics[scale=1]{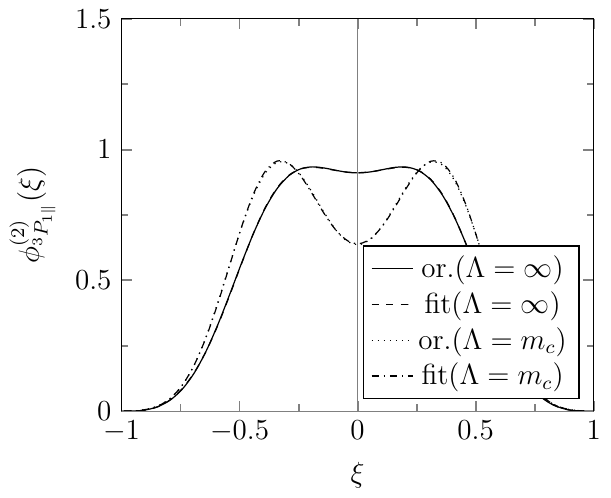} & \includegraphics[scale=1]{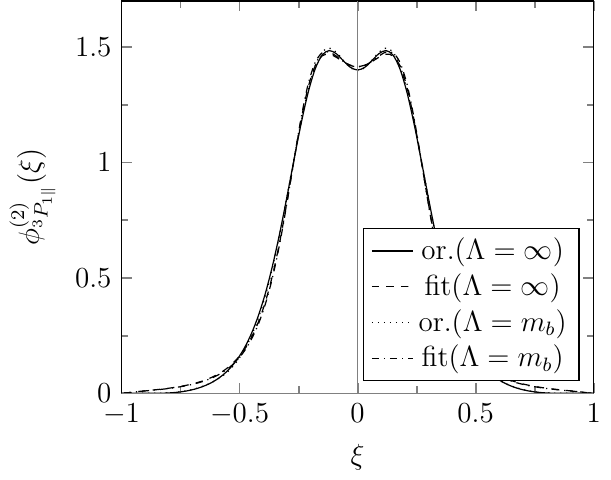} \\
\includegraphics[scale=1]{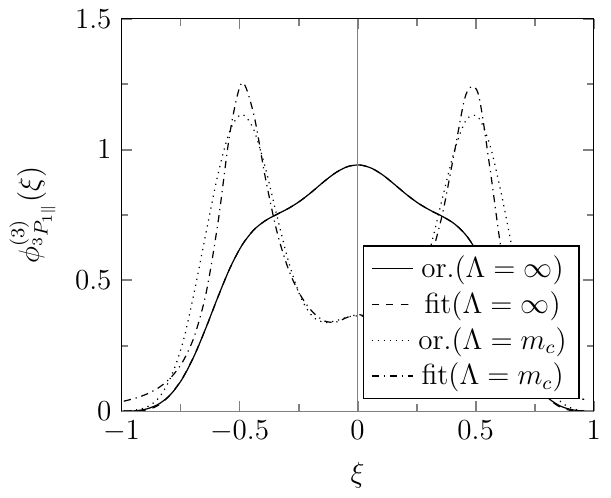} & \includegraphics[scale=1]{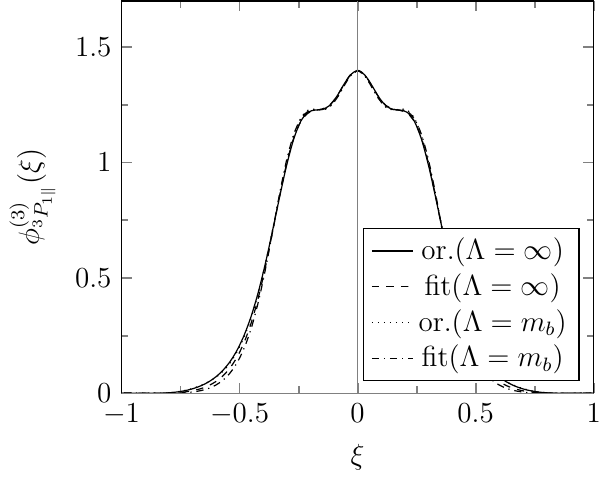} \\
\end{tabular}
\caption{LCDA plots as in Fig. (\ref{figilk}), but for $^{3} P_{1\parallel}$ states.}
\end{center}
\end{figure}
\newpage
\begin{figure}
\begin{center}
\begin{tabular}{cc}
$c\bar{c}$ & $b\bar{b}$ \\
\includegraphics[scale=1]{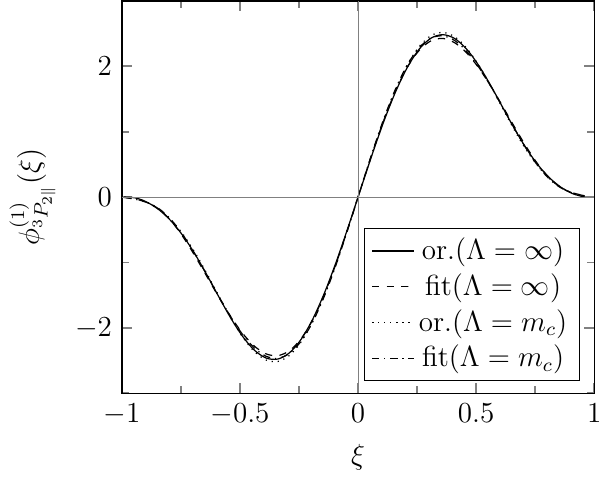} & \includegraphics[scale=1]{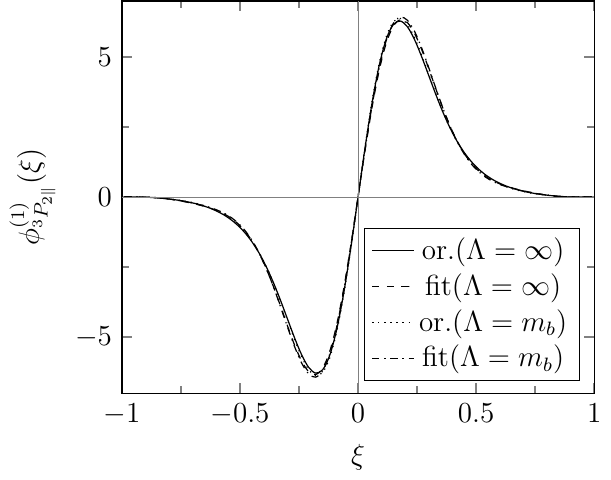} \\
\includegraphics[scale=1]{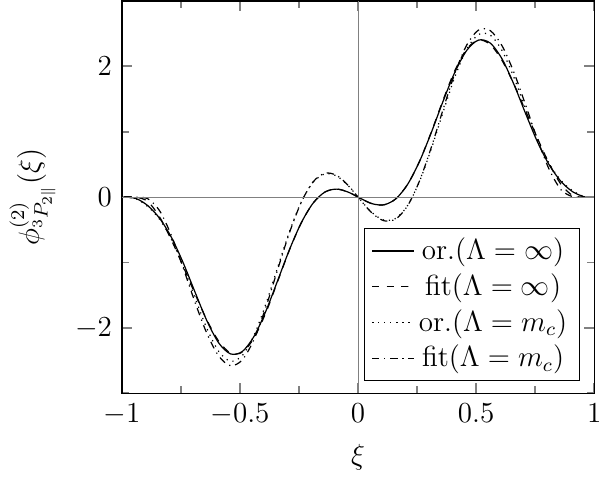} & \includegraphics[scale=1]{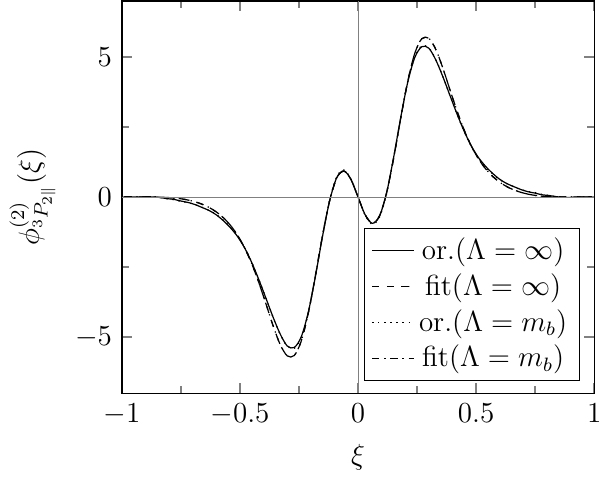} \\
\includegraphics[scale=1]{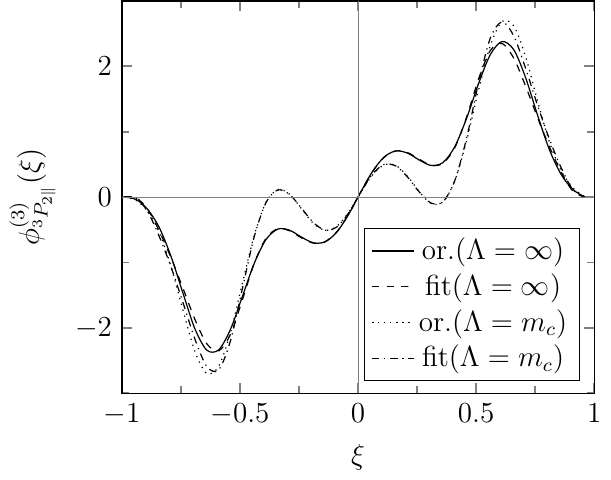} & \includegraphics[scale=1]{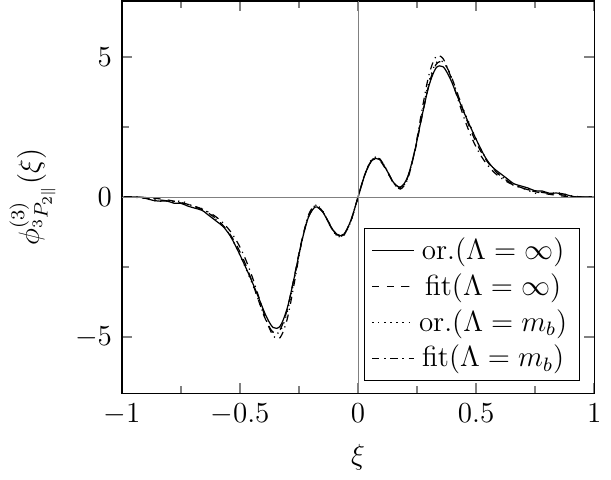} \\
\end{tabular}
\caption{LCDA plots as in Fig. (\ref{figilk}), but for $^{3} P_{2\parallel}$ states.}
\end{center}
\end{figure}
\newpage
\newpage
\begin{table}\centering
\begin{tabular}{|c|c|c|c|c|c|c|}
\hline
           &$a $            &$\xi _{0}$      &$\sigma ^{2}$       &$\beta $        &$b  $            & $c $              \tabularnewline
\hline
$n=1$ &$5.29486$  &$-$               &$-$                 &$0.867659$  &$1.36101$   & $0.352628$  \tabularnewline
\hline
$n=2$ &$6.98919$   &$0.257499$ &$0.182999$  &$1.50283$  &$-0.560786$  & $0.149576$   \tabularnewline
\hline
$n=3$ &$1.98475$  &$0.51822$   &$0.0928117$   &$0.847995$  &$0.482135$ & $0.083937$   \tabularnewline
\hline
\end{tabular}
\caption{$^{3}P_{0}$ charmonium fit parameters, $\Lambda =\infty $}
\label{table:6}
\end{table}

\begin{table}\centering
\begin{tabular}{|c|c|c|c|c|c|c|}
\hline
           &$a $            &$\xi _{0}$      &$\sigma ^{2}$       &$\beta $        &$b  $            & $c $              \tabularnewline
\hline
$n=1$ &$5.87696$  &$-$               &$-$                 &$1.00251$   &$1.2953$     & $0.346534$   \tabularnewline
\hline
$n=2$ &$7.26777$  &$0.255466$   &$0.185855$  &$1.52799$  &$-0.567534$  & $0.152166$   \tabularnewline
\hline
$n=3$ &$2.22227$  &$0.538532$  &$0.063798$   &$0.736003$  &$0.262874$ & $0.0490782$ \tabularnewline
\hline
\end{tabular}
\caption{$^{3}P_{0}$ charmonium fit parameters, $\Lambda =m_{c} $}
\end{table}

\begin{table}\centering
\begin{tabular}{|c|c|c|c|c|c|c|}
\hline
           &$a $            &$\xi _{0}$      &$\sigma ^{2}$       &$\beta $        &$b  $            & $c $              \tabularnewline
\hline
$n=1$ &$14.9131$  &$-$               &$-$                 &$1.29523$   &$2.47014$   & $0.070175$   \tabularnewline
\hline
$n=2$ &$4.08343$  &$0$ &$0.018278$ &$1.01745$ &$-0.155892$  & $0.0244758$ \tabularnewline
\hline
$n=3$ &$6.5$          &$0.288561$ &$0.0244281$   &$1.70944$   &$0.260454$ & $0.0285309$ \tabularnewline
\hline
\end{tabular}
\caption{$^{3}P_{0}$ bottomonium fit parameters, $\Lambda =\infty $}
\end{table}

\begin{table}\centering
\begin{tabular}{|c|c|c|c|c|c|c|}
\hline
           &$a $            &$\xi _{0}$      &$\sigma ^{2}$       &$\beta $        &$b  $            & $c $              \tabularnewline
\hline
$n=1$ &$7.88045$  &$-$               &$-$                 &$1.03344$    &$5.01843$   & $0.0756304$\tabularnewline
\hline
$n=2$ &$4.15287$      &$0$  &$0.0173548$ &$1.01512$    &$-0.173547$  & $0.0256279$ \tabularnewline
\hline
$n=3$ &$3.33759$  &$0.280277$  &$0.0230325$  &$1.10439$    &$0.283173$ & $0.0283235$ \tabularnewline
\hline
\end{tabular}
\caption{$^{3}P_{0}$ bottomonium fit parameters, $\Lambda =m_{b} $}
\end{table}

\begin{table}\centering
\begin{tabular}{|c|c|c|c|c|c|c|c|}
\hline
           &$a $            &$\xi _{0}$      &$\sigma ^{2}$       &$\beta $        &$b  $            & $c $              \tabularnewline
\hline
$n=1$ &$1.78365$  &$-$               &$-$                 &$1.4891$      &$0.636329$ & $0.431836$  \tabularnewline
\hline
$n=2$ &$2.43681$  &$0.284756$  &$0.0966684$  &$0.921282$  &$0$              & $-$                \tabularnewline
\hline
$n=3$ &$1.53215$  &$0.487944$  &$0.122605$    &$0.734715$  &$0.562724$ & $0.122482$   \tabularnewline
\hline
\end{tabular}
\caption{$^{1}P_{1\perp }$ charmonium fit parameters, $\Lambda =\infty $}
\end{table}

\begin{table}\centering
\begin{tabular}{|c|c|c|c|c|c|c|c|}
\hline
           &$a $            &$\xi _{0}$      &$\sigma ^{2}$       &$\beta $        &$b  $            & $c $              \tabularnewline
\hline
$n=1$ &$2.26917$  &$-$               &$-$                 &$2.14573$   &$0.541302$ & $0.383543$   \tabularnewline
\hline
$n=2$ &$1.40861$ &$0.370795$  &$0.0301022$   &$0.303023$ &$0$               & $-$                \tabularnewline
\hline
$n=3$ &$2.15625$  &$0.530031$  &$0.0483686$  &$0.23274$   &$0.269402$  & $1.01888$    \tabularnewline
\hline
\end{tabular}
\caption{$^{1}P_{1\perp }$ charmonium fit parameters, $\Lambda =m_{c} $}
\end{table}
\clearpage

\begin{table}\centering
\begin{tabular}{|c|c|c|c|c|c|c|c|}
\hline
           &$a $            &$\xi _{0}$      &$\sigma ^{2}$       &$\beta $        &$b  $            & $c $              \tabularnewline
\hline
$n=1$ &$11.7196$  &$-$               &$-$                 &$1.74248$    &$0.126342$ & $0.0638176$ \tabularnewline
\hline
$n=2$ &$1.59483$  &$0.12902$   &$0.00748694$ &$0.1$            &$0$              & $-$                 \tabularnewline
\hline
$n=3$ &$17.2954$  &$0.254288$ &$0.0186876$   &$2.57763$    &$0.18766$   & $0.0149827$ \tabularnewline
\hline
\end{tabular}
\caption{$^{1}P_{1\perp }$ bottomonium fit parameters, $\Lambda =\infty $}
\end{table}

\begin{table}\centering
\begin{tabular}{|c|c|c|c|c|c|c|c|}
\hline
           &$a $            &$\xi _{0}$      &$\sigma ^{2}$       &$\beta $        &$b  $            & $c $              \tabularnewline
\hline
$n=1$ &$13.3807$  &$-$               &$-$                 &$1.94856$   &$0.120252$  & $0.0635912$\tabularnewline
\hline
$n=2$ &$1.61774$   &$0.135557$ &$0.00687409$&$0.1$           &$0$               & $-$                \tabularnewline
\hline
$n=3$ &$8.28066$  &$0.24319$    &$0.0150691$  &$1.86152$   &$0.242844$  & $0.0159363$ \tabularnewline
\hline
\end{tabular}
\caption{$^{1}P_{1\perp }$ bottomonium fit parameters, $\Lambda =m_{b} $}
\end{table}

\begin{table}\centering
\begin{tabular}{|c|c|c|c|c|c|c|}
\hline
           &$a $            &$\xi _{0}$      &$\sigma ^{2}$       &$\beta $        &$b  $            & $c $              \tabularnewline
\hline
$n=1$ &$7.28898$  &$-$               &$-$                 &$4.6828$     &$1.79244$   & $0.273028$  \tabularnewline
\hline
$n=2$ &$8.65428$  &$0.242683$ &$0.167489$   &$1.86131$ &$-0.326097$    & $0.128336$   \tabularnewline
\hline
$n=3$ &$2.8194$    &$0.514286$ &$0.0924345$   &$1.13424$   &$0.456194$ & $0.0903501$   \tabularnewline
\hline
\end{tabular}
\caption{$^{1}P_{1\parallel }$ charmonium fit parameters, $\Lambda =\infty $}
\end{table}

\begin{table}\centering
\begin{tabular}{|c|c|c|c|c|c|c|}
\hline
           &$a $            &$\xi _{0}$      &$\sigma ^{2}$       &$\beta $        &$b  $            & $c $              \tabularnewline
\hline
$n=1$ &$53.6273$  &$-$               &$-$                 &$2.39208$   &$0.155202$ & $0.273479$  \tabularnewline
\hline
$n=2$ &$10.0745$  &$0.269738$ &$0.160284$  &$1.92211$    &$-0.426175$  & $0.13256$   \tabularnewline
\hline
$n=3$ &$2.74678$  &$0.516953$  &$0.0704437$  &$0.980324$ &$0.294645$ & $0.0616262$ \tabularnewline
\hline
\end{tabular}
\caption{$^{1}P_{1\parallel }$ charmonium fit parameters, $\Lambda =m_{c} $}
\end{table}
 
\begin{table}\centering
\begin{tabular}{|c|c|c|c|c|c|c|}
\hline
      &$a $            &$\xi _{0}$      &$\sigma ^{2}$   &$\beta $   &$b  $    & $c $   \tabularnewline
\hline
$n=1$ &$10.5205$  &$-$               &$-$                 &$1.50861$   &$5.65674$ &$0.0638543$\tabularnewline
\hline
$n=2$ &$24.2512$  &$0$  &$0.0147382$  &$2.63148$   &$-0.199554$   & $0.0189217$\tabularnewline
\hline
$n=3$ &$3.28997$  &$0.232939$  &$0.0123631$&$0.987534$   &$0.300752$  & $0.0169844$ \tabularnewline
\hline
\end{tabular}
\caption{$^{1}P_{1\parallel }$ bottomonium fit parameters, $\Lambda =\infty $}
\end{table}

\clearpage
\begin{table}\centering
\begin{tabular}{|c|c|c|c|c|c|c|}
\hline
           &$a $            &$\xi _{0}$      &$\sigma^{2} $       &$\beta $        &$b  $            & $c $              \tabularnewline
\hline
$n=1$ &$9.97137$  &$-$               &$-$                 &$1.52337$   &$6.0868$     & $0.0639454$\tabularnewline
\hline
$n=2$ &$8.82176$  &$1.25\times 10^{-6}$  &$0.0113222$  &$1.63918$   &$-0.18903$  & $0.0186983$ \tabularnewline
\hline
$n=3$ &$18.256$    &$0.255062$  &$0.0130517$  &$2.61225$   &$0.240445$ & $0.0151002$ \tabularnewline
\hline
\end{tabular}
\caption{$^{1}P_{1\parallel }$ bottomonium fit parameters, $\Lambda =m_{b} $}
\end{table}

\begin{table}\centering
\begin{tabular}{|c|c|c|c|c|c|c|}
\hline
           &$a $            &$\xi _{0}$      &$\sigma ^{2}$       &$\beta $        &$b  $            & $c $              \tabularnewline
\hline
$n=1$ &$47.933$    &$-$               &$-$                 &$2.33766$    &$0.183997$& $0.263943$   \tabularnewline
\hline
$n=2$ &$9.78918$  &$0.142262$    &$0.180577$  &$1.92684$  &$-0.391433$ & $0.12737$ \tabularnewline
\hline
$n=3$ &$2.62768$  &$0.474425$  &$0.07959$     &$1.05711$     &$0.437048$ & $0.0611108$ \tabularnewline
\hline
\end{tabular}
\caption{$^{3}P_{1\perp }$ charmonium fit parameters, $\Lambda =\infty $}
\end{table}

\begin{table}\centering
\begin{tabular}{|c|c|c|c|c|c|c|}
\hline
           &$a $            &$\xi _{0}$      &$\sigma ^{2}$       &$\beta $        &$b  $            & $c $              \tabularnewline
\hline
$n=1$ &$58.1733$   &$-$               &$-$                 &$2.42818$   &$0.148617$& $0.258012$   \tabularnewline
\hline
$n=2$ &$12.3322$  &$0.156543$  &$0.184174$   &$2.02338$  &$-0.500837$  & $0.135897$  \tabularnewline
\hline
$n=3$ &$2.71267$  &$0.487316$  &$0.062559$   &$0.966985$  &$0.318898$ & $0.044592$  \tabularnewline
\hline
\end{tabular}
\caption{$^{3}P_{1\perp }$ charmonium fit parameters, $\Lambda =m_{c} $}
\end{table}

\begin{table}\centering
\begin{tabular}{|c|c|c|c|c|c|c|}
\hline
           &$a $            &$\xi _{0}$      &$\sigma ^{2}$       &$\beta $        &$b  $            & $c $              \tabularnewline
\hline
$n=1$ &$460.545$  &$-$               &$-$                 &$3.96925$   &$0.127859$  & $0.0535103$\tabularnewline
\hline
$n=2$ &$24.2512$  &$0$    &$0.0147382$     &$2.63148$  &$-0.199554$  & $0.0189217$    \tabularnewline
\hline
$n=3$ &$1.33436$  &$0.207932$ &$0.009408$     &$0.1$          &$0.314583$   & $0.0117902$ \tabularnewline
\hline
\end{tabular}
\caption{$^{3}P_{1\perp }$ bottomonium fit parameters, $\Lambda =\infty $}
\end{table}

\begin{table}\centering
\begin{tabular}{|c|c|c|c|c|c|c|}
\hline
           &$a $            &$\xi _{0}$      &$\sigma ^{2}$       &$\beta $        &$b  $            & $c $              \tabularnewline
\hline
$n=1$ &$460.778$  &$-$               &$-$                 &$3.98292$   &$0.129095$ & $0.515475$  \tabularnewline
\hline
$n=2$ &$24.2512$ &$0$   &$0.0147382$  &$2.63148$ &$-0.199554$  & $0.0189217$    \tabularnewline
\hline
$n=3$ &$2.50176$ &$0.21579$    &$0.00891723$ &$0.692022$ &$0.312109$  & $0.0119042$\tabularnewline
\hline
\end{tabular}
\caption{$^{3}P_{1\perp }$ bottomonium fit parameters, $\Lambda =m_{b} $}
\end{table}

\begin{table}\centering
\begin{tabular}{|c|c|c|c|c|c|c|c|}
\hline
           &$a $            &$\xi _{0}$      &$\sigma ^{2}$       &$\beta $        &$b  $            & $c $              \tabularnewline
\hline
$n=1$ &$1.89725$  &$-$               &$-$                 &$0.833424$ &$0.531147$  & $0.334782$  \tabularnewline
\hline
$n=2$ &$3.59963$  &$0.000054$  &$0.148262$    &$0.571469$ &$-0.551464$ & $0.67177$    \tabularnewline
\hline
$n=3$ &$2.32785$  &$0.469035$  &$0.109016$    &$0.999494$ &$0.406331$  & $0.0780892$ \tabularnewline
\hline
\end{tabular}
\caption{$^{3}P_{1\parallel }$ charmonium fit parameters, $\Lambda =\infty $}
\end{table}

\begin{table}\centering
\begin{tabular}{|c|c|c|c|c|c|c|c|}
\hline
           &$a $            &$\xi _{0}$      &$\sigma ^{2}$       &$\beta $        &$b  $            & $c $              \tabularnewline
\hline
$n=1$ &$11.121$    &$-$               &$-$                 &$1.91255$    &$0.0836092$& $0.343286$ \tabularnewline
\hline
$n=2$ &$5.84685$ &$1\times 10^{-8}$&$0.152987$&$0.33432$  &$-0.847095$& $0.529515$ \tabularnewline
\hline
$n=3$ &$1.34405$ &$0.487261$    &$0.0170547$   &$0.05$       &$0.0564819$& $0.00851709$\tabularnewline
\hline
\end{tabular}
\caption{$^{3}P_{1\parallel }$ charmonium fit parameters, $\Lambda =m_{c} $}
\end{table}

\begin{table}\centering
\begin{tabular}{|c|c|c|c|c|c|c|c|}
\hline
           &$a $            &$\xi _{0}$      &$\sigma ^{2}$       &$\beta $        &$b  $            & $c $              \tabularnewline
\hline
$n=1$ &$15.1067$  &$-$               &$-$                 &$1.99935$   &$0.104585$ & $0.0630898$ \tabularnewline
\hline
$n=2$ &$1.62749$  &$0.129254$  &$0.00695058$&$0.1$           &$0$              & $-$                 \tabularnewline
\hline
$n=3$ &$4.52742$  &$0.216833$  &$0.0131697$  &$1.26096$   &$0.233993$ & $0.0111891$  \tabularnewline
\hline
\end{tabular}
\caption{$^{3}P_{1\parallel }$ bottomonium fit parameters, $\Lambda =\infty $}
\end{table}

\begin{table}\centering
\begin{tabular}{|c|c|c|c|c|c|c|c|}
\hline
           &$a $            &$\xi _{0}$      &$\sigma ^{2}$       &$\beta $        &$b  $            & $c $              \tabularnewline
\hline
$n=1$ &$8.43243$  &$-$               &$-$                 &$1.68704$    &$0.200742$ & $0.0638126$\tabularnewline
\hline
$n=2$ &$1.63952$  &$0.133876$  &$0.00654186$&$0.1$            &$0$              & $-$                \tabularnewline
\hline
$n=3$ &$33.7026$  &$0.252602$  &$0.0148385$  &$3.14863$   &$0.181075$  & $0.00958596$\tabularnewline
\hline
\end{tabular}
\caption{$^{3}P_{1\parallel }$ bottomonium fit parameters, $\Lambda =m_{b} $}
\end{table}

\begin{table}\centering
\begin{tabular}{|c|c|c|c|c|c|c|}
\hline
           &$a $            &$\xi _{0}$      &$\sigma ^{2}$       &$\beta $        &$b  $            & $c $              \tabularnewline
\hline
$n=1$ &$18.9386$  &$-$               &$-$                 &$1.53348$   &$0.588362$ & $0.187505$  \tabularnewline
\hline
$n=2$ &$9.78268$  &$0.000301$  &$0.159954$   &$1.92414$    &$-0.331021$  & $0.125204$   \tabularnewline
\hline
$n=3$ &$2.98352$  &$0.455036$  &$0.0789049$ &$1.21573$    &$0.49737$   & $0.0706649$ \tabularnewline
\hline
\end{tabular}
\caption{$^{3}P_{2\perp }$ charmonium fit parameters, $\Lambda =\infty $}
\end{table}

\begin{table}\centering
\begin{tabular}{|c|c|c|c|c|c|c|}
\hline
           &$a $            &$\xi _{0}$      &$\sigma ^{2}$       &$\beta $        &$b  $            & $c $              \tabularnewline
\hline
$n=1$ &$23.5577$  &$-$               &$-$                 &$1.70536$    &$0.469824$ & $0.191909$  \tabularnewline
\hline
$n=2$ &$6.95313$  &$0.2615$    &$0.122794$  &$1.72871$    &$-0.251262$  & $0.0784634$   \tabularnewline
\hline
$n=3$ &$3.28322$  &$0.4731$      &$0.0684879$ &$1.16459$    &$0.309103$  & $0.0461067$ \tabularnewline
\hline
\end{tabular}
\caption{$^{3}P_{2\perp }$ charmonium fit parameters, $\Lambda =m_{c} $}
\end{table}

\begin{table}\centering
\begin{tabular}{|c|c|c|c|c|c|c|}
\hline
           &$a $            &$\xi _{0}$      &$\sigma ^{2}$       &$\beta $        &$b  $            & $c $              \tabularnewline
\hline
$n=1$ &$5.03627$  &$-$               &$-$                 &$1.36603$   &$12.6441$   & $0.066995$   \tabularnewline
\hline
$n=2$ &$54.0733$  &$0$  &$0.0147382$  &$3.38888$   &$-0.199327$  & $0.0189271$ \tabularnewline
\hline
$n=3$ &$4.27052$    &$0.223597$  &$0.0109257$&$1.21376$  &$0.305811$ & $0.0155296$ \tabularnewline
\hline
\end{tabular}
\caption{$^{3}P_{2\perp }$ bottomonium fit parameters, $\Lambda =\infty $}
\end{table}

\begin{table}\centering
\begin{tabular}{|c|c|c|c|c|c|c|}
\hline
           &$a $            &$\xi _{0}$      &$\sigma ^{2}$       &$\beta $        &$b  $            & $c $              \tabularnewline
\hline
$n=1$ &$4.60379$  &$-$               &$-$                 &$1.34201$   &$14.0572$   & $0.0666158$ \tabularnewline
\hline
$n=2$ &$22.905$  &$1.04\times 10^{-6}$ &$0.0121405$  &$2.55386$    &$-0.184869$  & $0.0175767$\tabularnewline
\hline
$n=3$ &$7.66472$  &$0.231103$  &$0.0112419$  &$1.75192$   &$0.282943$  & $0.014799$  \tabularnewline
\hline
\end{tabular}
\caption{$^{3}P_{2\perp }$ bottomonium fit parameters, $\Lambda =m_{b} $}
\end{table}

\begin{table}\centering
\begin{tabular}{|c|c|c|c|c|c|c|}
\hline
           &$a $            &$\xi _{0}$      &$\sigma ^{2}$       &$\beta $        &$b  $            & $c $              \tabularnewline
\hline
$n=1$ &$11.3608$  &$-$               &$-$                 &$1.21471$    &$0.697849$ & $0.272678$  \tabularnewline
\hline
$n=2$ &$4.44771$  &$0.232039$  &$0.158319$    &$1.41475$  &$-0.119605$  & $0.0808185$  \tabularnewline
\hline
$n=3$ &$2.3202$    &$0.461568$  &$0.0998141$  &$1.01046$    &$0.418644$ & $0.0858612$\tabularnewline
\hline
\end{tabular}
\caption{$^{3}P_{2\parallel }$ charmonium fit parameters, $\Lambda =\infty $}
\end{table}

\begin{table}\centering
\begin{tabular}{|c|c|c|c|c|c|c|}
\hline
           &$a $            &$\xi _{0}$      &$\sigma ^{2}$       &$\beta $        &$b  $            & $c $              \tabularnewline
\hline
$n=1$ &$14.4417$  &$-$               &$-$                 &$1.40581$    &$0.551405$ & $0.278178$  \tabularnewline
\hline
$n=2$ &$8.83337$  &$0.173308$  &$0.229571$    &$1.85411$  &$-0.360409$  & $0.141981$      \tabularnewline
\hline
$n=3$ &$2.51931$  &$0.476137$  &$0.0902674$  &$0.976052$  &$0.253069$ & $0.0561185$\tabularnewline
\hline
\end{tabular}
\caption{$^{3}P_{2\parallel }$ charmonium fit parameters, $\Lambda =m_{c} $}
\end{table}

\begin{table}\centering
\begin{tabular}{|c|c|c|c|c|c|c|}
\hline
           &$a $            &$\xi _{0}$      &$\sigma ^{2}$       &$\beta $        &$b  $            & $c $              \tabularnewline
\hline
$n=1$ &$5.24954$  &$-$               &$-$                &$1.11938$     &$10.5179$   & $0.069816$  \tabularnewline
\hline
$n=2$ &$26.8594$  &$0$  &$0.0147382$ &$2.73992$    &$-0.168346$  & $0.0189217$ \tabularnewline
\hline
$n=3$ &$3.78742$  &$0.222412$  &$0.0131263$  &$1.13605$  &$0.259637$  & $0.0157952$   \tabularnewline
\hline
\end{tabular}
\caption{$^{3}P_{2\parallel }$ bottomonium fit parameters, $\Lambda =\infty $}
\end{table}

\begin{table}\centering
\begin{tabular}{|c|c|c|c|c|c|c|}
\hline
           &$a $            &$\xi _{0}$      &$\sigma ^{2}$       &$\beta $        &$b  $            & $c $              \tabularnewline
\hline
$n=1$ &$4.7464$    &$-$               &$-$                 &$1.09225$    &$11.8753$   & $0.0695242$\tabularnewline
\hline
$n=2$ &$26.8594$    &$0$  &$0.0147382$ &$2.73992$    &$-0.168346$  & $0.0189217$\tabularnewline
\hline
$n=3$ &$2.28$        &$0.222311$  &$0.0105893$  &$0.673188$  &$0.326787$ & $0.0180464$ \tabularnewline
\hline
\end{tabular}
\caption{$^{3}P_{2\parallel }$ bottomonium fit parameters, $\Lambda =m_{b} $}
\label{table:33}
\end{table}

\end{document}